\documentclass[]{aa}  
\pdfoutput=1
\usepackage{graphicx}
\usepackage{caption}
\usepackage{subcaption}
\usepackage{epsf}
\usepackage{longtable,lscape}
\usepackage{dcolumn}
\usepackage{booktabs} 
\usepackage{url}
\usepackage{float}
\usepackage{microtype}
\usepackage[varg]{txfonts}
\usepackage{natbib,twoopt} 
\usepackage{longtable}
\usepackage{lipsum}
\def\apjl{ApJ}%
\def\apjs{ApJS}%
\def\ao{Appl.~Opt.}%
\def\aap{A\&A}%
\usepackage{xcolor}
\bibpunct{(}{)}{;}{a}{}{,} 

\let\oldhat\hat

\renewcommand{\hat}[1]{\oldhat{\mathbf{#1}}}
\newcommand{\kms}{km\,${\rm s}^{-1}$}

\newcommand{\gcgs}{$[\text{cm}\,\text{s}^{-2}]$\,}

\newcommand{\Lsun}{\mbox{$L_\odot$}}

\newcommand{\myr}{\mbox{$M_\odot\,{\rm yr}^{-1}$}}

\newcommand{\lum}{erg\,s$^{-1}$}

\titlerunning{SMC\,AB\,6 is a high-order multiple system}

\begin{document}

   \title{The Shortest-period Wolf-Rayet binary in the Small Magellanic Cloud: Part of a high-order multiple system}

   \subtitle{Spectral and orbital analysis of SMC\,AB\,6}

   \author{T.\ Shenar\inst{1} 
          \and R.\ Hainich\inst{1}
          \and H.\ Todt\inst{1}
          \and A.\ F.\ J.\ Moffat\inst{2}     
          \and A.\ Sander\inst{1}       
          \and L.\ M.\ Oskinova\inst{1}            
          \and V.\ Ramachandran\inst{1}   
          \and M.\ Munoz\inst{3}          
          \and H.\ Pablo\inst{5}                    
          \and H.\ Sana\inst{4}          
          \and W.-R.\ Hamann\inst{1}         
          }
          
   \institute{\inst{1}{Institut f\"ur Physik und Astronomie, Universit\"at Potsdam,
                Karl-Liebknecht-Str. 24/25, D-14476 Potsdam, Germany}\\
              \email{shtomer@astro.physik.uni-potsdam.de}  \\                
              \inst{2}{D\'epartement de physique and Centre de Recherche en Astrophysique 
                du Qu\'ebec (CRAQ), Universit\'e de Montr\'eal, C.P. 6128, Succ.~Centre-Ville, Montr\'eal, Qu\'ebec, H3C 3J7, Canada}\\        
              \inst{3}{Queen’s University 99 University Ave, Kingston, Ontario, Canada} \\      
              \inst{4}{Institute of Astrophysics, KU Leuven, Celestijnenlaan 200 D, 3001, Leuven, Belgium} \\        
              \inst{5}{American Association of Variable Star Observers, 49 Bay State Road, Cambridge, MA 02138, USA} \\                     
              }
   \date{Received ? / Accepted ?}


\abstract
{SMC AB\,6 is the shortest-period ($P{=}6.5\,$d) Wolf-Rayet (WR) binary in the Small Magellanic Cloud. This binary is therefore 
a key system in the study of binary interaction and formation of WR stars at low metallicity. The WR component in 
AB\,6 was previously found to be very luminous ($\log L{=}6.3\,[\Lsun]$) compared to 
its reported orbital mass (${\approx}8\,M_\odot$), placing it significantly above the Eddington limit.
}
{
Through spectroscopy and orbital analysis of newly acquired optical data taken with the Ultraviolet and Visual Echelle Spectrograph (UVES), 
we aim to understand the peculiar results reported for this system 
and explore its evolutionary history.  
}
{We measured radial velocities via cross-correlation and performed a spectral analysis using the Potsdam Wolf-Rayet 
model atmosphere code. The evolution of the system was analyzed using the Binary Population and Spectral Synthesis evolution code.}
{AB\,6 contains at least four stars. 
The 6.5\,d period WR binary comprises the WR primary (WN3:h, star A) and 
a rather rapidly rotating ($v_\text{eq}{=}265\,$\kms) early O-type companion (O5.5~V, star B). 
Static N\,{\sc iii} and N\,{\sc iv} emission lines and absorption signatures in He lines 
suggest the presence of an early-type emission line star (O5.5~I(f), star C). Finally, narrow absorption lines portraying a long-term radial velocity variation show 
the existence of a fourth star (O7.5~V, star D). Star D appears to form a second $140\,$d period binary together with 
a fifth stellar member, which is a B-type dwarf or a black hole. It is not clear that these additional components are bound to the WR binary. We derive a mass ratio 
of $M_\text{O}/M_\text{WR}{=}2.2\pm0.1$. 
The 
WR star is found to be less luminous than previously thought ($\log L{=}5.9\,[L_\odot]$) and, adopting $M_\text{O}{=}41\,M_\odot$ for star B,
more massive ($M_\text{WR}{=}18\,M_\odot$).
Correspondingly, the WR star does not exceed the 
Eddington limit. We derive the initial masses of $M_\text{i,WR}{=}60\,M_\odot$ and $M_\text{i,O}{=}40\,M_\odot$ and an age of $3.9\,$Myr for the system. The WR binary 
likely experienced nonconservative mass transfer in the past supported by the relatively rapid rotation of star B.
}
{
Our study shows that AB\,6 is a multiple--probably quintuple--system. This finding resolves the previously reported puzzle of the WR primary exceeding the Eddington limit 
and suggests that the WR star exchanged mass with its companion in the past.
}
\keywords{Stars: massive -- Binaries: spectroscopic --  Stars: Wolf-Rayet -- Magellanic Clouds   -- Stars: individual: SMC\,AB\,6 -- Stars: atmospheres}

\maketitle

\section{Introduction}
\label{sec:introduction}

The study of massive stars ($M_\text{i} \gtrsim 8\,M_\odot$) and binaries at various metallicities 
is essential for a multitude of astrophysical fields, from supernovae physics to galactic evolution \citep[e.g.,][]{Langer2012}.
Stars that are massive enough eventually reach the classical  
Wolf-Rayet (WR) phase that is characterized by powerful stellar winds and hydrogen depletion. 
Studying WR stars is important both 
for understanding the evolution of massive stars \citep[e.g.,][]{Crowther2007} and for constraining the energy budget of galaxies \citep{Ramachandran2018, Ramachandran2018b}.
It is essential to improve our understanding of WR stars especially in the era of gravitational waves .

Two WR-star formation channels have been proposed. First, single massive stars  can lose their hydrogen-rich envelopes via 
powerful radiation-driven winds or eruptions \citep{Conti1976}.
Second, mass donors in binary systems, which are expected to be common \citep[e.g.,][]{Kiminki2012, Sana2012, Almeida2017}, 
may lose their outer layers through mass transfer \citep{Paczynski1973, Vanbeveren1998b}. 
Since wind mass-loss rates $\dot{M}$ scale with metallicity $Z$ \citep{Crowther2006, Hainich2015}, it is  
expected that the binary formation channel should become dominant in low metallicity environments \citep{Maeder1994, Bartzakos2001}. 
The Small Magellanic Cloud (SMC), with $Z_\text{SMC}{\approx} 1/7\,Z_\odot$ \citep{Trundle2007} and a distance of $62\,$kpc \citep{Keller2006}, offers an 
ideal environment to test this. The reported SMC WR binary fraction is 
similar to the Galactic  WR binary fraction \citep{Foellmi2003SMC} and 
the dominance of the binary channel in the SMC is still debated \citep{Shenar2016, Schootemeijer2017}.

\begin{figure}[]
\centering
  \includegraphics[width=\columnwidth]{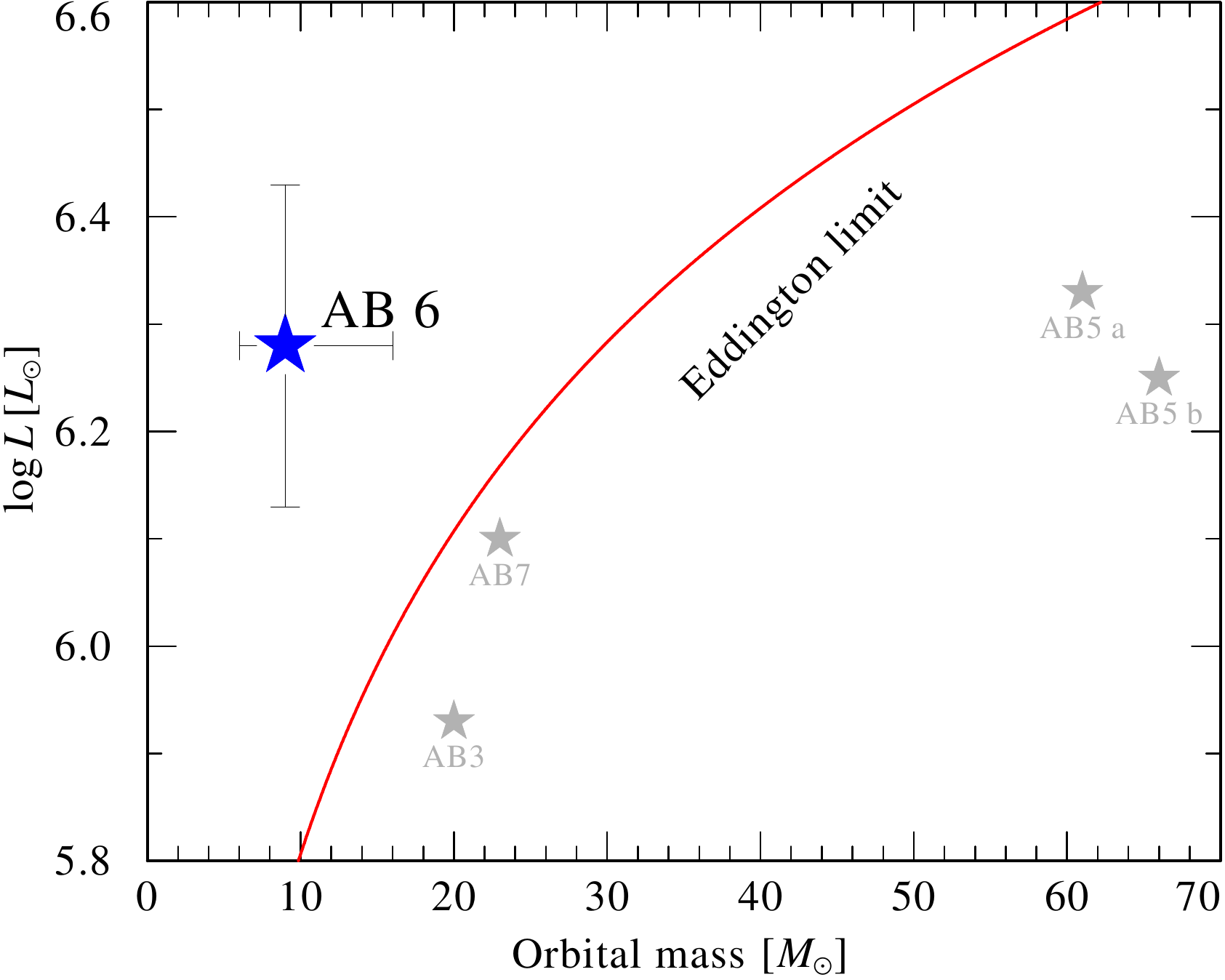}
  \caption{Luminosity and orbital mass originally derived for the WR component of AB\,6 by \citet{Shenar2016}, 
  compared to the Eddington limit for a fully ionized helium atmosphere. 
  The positions of the other SMC WN stars in binaries are also shown. The values for AB\,6 are revised in the present paper.}
\label{fig:Edin}
\end{figure}

Our target, \object{SMC\,AB\,6} (AB\,6 hereafter), has the shortest period among the five known SMC WR binaries, and hence offers 
a unique laboratory to investigate the origin of WR stars at low metallicity. \cite{Azzopardi1979}
originally classified it as a binary (WN3 + O7 Ia) based on the strong dilution of the WR  emission lines 
and the presence of absorption 
features in its optical spectrum. \citet{Moffat1982} later proved its binarity based on radial velocity (RV) measurements.
The latest orbital study of the system was performed by \citet{Foellmi2003SMC}, who classified AB\,6 as WN4:${+}$O6.5~I and derived 
a period of $P=6.5$\,d. Assuming an inclination of $i{=}65^\circ$, the orbital 
parameters derived by \citet{Foellmi2003SMC} imply a mass of $M_\text{WR}{=}7.5\,M_\odot$, which is comparable with the findings 
of \citet{Hutchings1984}. 
Despite its relatively low Keplerian mass, the WR primary was reported to have a very high luminosity of $\log L{=}6.3\,[L_\odot$] by 
\citet{Shenar2016}, who analyzed optical and UV spectra of the system. This result placed the WR star significantly 
above the Eddington limit (Fig.\,\ref{fig:Edin}). Super-Eddington stars should be unstable on a dynamical timescale, portray 
strong eruptions, and generally be strongly variable \citep{Shaviv2001}, none of which are observed for AB\,6. 

These peculiar results encouraged us to acquire 
high quality UVES spectra of AB\,6. The main goal 
of this study is to explore whether spectra of unprecedented quality can be used to resolve the Eddington limit problem 
and to investigate the evolutionary history of the shortest-period SMC WR binary.
The observations used in this study are described in Sect.\,\ref{sec:obsdata}. In Sect.\,\ref{sec:binquad}, we illustrate how 
phase resolved UVES observations reveal the presence of additional components in the spectrum of AB\,6. In Sect.\,\ref{sec:orban}, we perform an orbital 
analysis of AB\,6, while in Sect.\,\ref{sec:specan} we present a spectral analysis of AB\,6. The nature of the system 
in light of our results is discussed in Sect.\,\ref{sec:disc}, and a summary is given in Sect.\,\ref{sec:summary}.

\section{Observational data}
\label{sec:obsdata}

Our study relies predominantly on a spectroscopic dataset obtained with the UVES spectrograph at ESO's Very Large Telescope
in 2017 (ID: 099.D-0766(A), P.I.: Shenar). Altogether 54 individual spectra were obtained with a slit width and a nominal seeing of 1.4'' 
during nine nights throughout period\,99. The observations were scheduled to allow for an almost
evenly spaced coverage of the orbit with $\Delta \phi{\approx}0.1$. 
During each night, six exposures covering the blue ($3730{-}5000\,$\AA) and red ($5655{-}9464$\,\AA) spectral bands were secured. The spectra were co-added 
for each of the nine distinct phases and were calibrated to a heliocentric frame of reference. 
A log of the co-added spectra can be found in Table\,\ref{tab:RVs}. 
Since we do not require an extremely precise wavelength calibration, we relied on the default data reduction provided by the ESO archive. 
The standard setting DIC-2\,437+760 was used. With a total exposure time of 50 minutes per night, a signal-to-noise 
ratio (S/N) of ${\approx} 100{-}130$ in the blue band and ${\approx}80{-}100$ in the red band at a resolving power of $R{\approx}30\,000$ was obtained. 
Phases given below correspond to the ephemeris in Table\,\ref{tab:orbpar}.

\renewcommand{\arraystretch}{1.2}
\setlength{\tabcolsep}{1.2mm}
\begin{table}
\small
\caption{Log of UVES observations and derived RVs for AB\,6}
\begin{center}
\begin{tabular}{c  c  c c  c  c c}
\hline
MJD\tablefootmark{a}       & $\phi$\tablefootmark{b} & $\text{RV}_\text{A,N\,{\sc v}}$\tablefootmark{c}    & $\text{RV}_\text{A, He\,{\sc ii}}$\tablefootmark{d}       & $\text{RV}_\text{B}$\tablefootmark{e} & $\text{RV}_\text{C}$\tablefootmark{f} & $\text{RV}_\text{D}$\tablefootmark{g} \\ 
\                          &                          &            [\kms]                        &            [\kms]                             & [\kms]                                       & [\kms]                                       &  [\kms]                                        \\ 
\hline
57938.30                  & 0.82                    & $19{\pm}10$                              & $32{\pm}25$                                  & $297{\pm}10$                                             & $211{\pm}15$                                   &  $143{\pm}3$ \\ 
57944.35                  & 0.74                    & $-8 {\pm}10$                             & $-55 {\pm}25$                                & $294{\pm}10$                                         & $205{\pm}15$                                    & $142{\pm}3$ \\ 
57945.38                  & 0.90                    & $78 {\pm}10$                             & $184 {\pm}25$                                 & $263{\pm}15$                                         & $213{\pm}10$                                    & $141{\pm}3$ \\ 
57947.33                  & 0.20                    & $457{\pm}10$                             & $428{\pm}25$                                 & $94{\pm}10$                                         & $218{\pm}10$                                    & $141{\pm}3$ \\ 
57959.34                  & 0.03                    & $295{\pm}10$                             & $328{\pm}25$                                  & $168{\pm}15$                                         & $214{\pm}10$                                    & $146{\pm}3$ \\ 
57963.30                  & 0.64                    & $36 {\pm}10$                             & $-61{\pm}25$                                  & $291{\pm}10$                                         & $213{\pm}10$                                    & $148{\pm}3$ \\ 
57974.32                  & 0.33                    & $437{\pm}10$                             & $377{\pm}25$                                  & $101{\pm}10$                                         & $214{\pm}10$                                    & $158{\pm}3$ \\ 
58001.23                  & 0.44                    & $336{\pm}10$                             & $172{\pm}25$                                  & $159{\pm}15$                                         & $200{\pm}10$                                    & $178{\pm}3$ \\ 
58025.16                  & 0.10                    & $380{\pm}10$                             & $464{\pm}25$                                  & $127{\pm}10$                                         & $216{\pm}10$                                    & $205{\pm}3$ \\ 
\hline                    
\end{tabular}
\tablefoot{
\tablefoottext{a}{Mid-exposure, in JD${-}$2\,400\,000.5}
\tablefoottext{b}{Using ephemeris in Table\,\ref{tab:orbpar}.}
\tablefoottext{c}{Using the N\,{\sc v}\,$\lambda \lambda 4604, 4620$ doublet. A shift of $-35$\,\kms\, should be applied for the estimated absolute RVs 
(Sect.\,\ref{sec:orban}).}
\tablefoottext{d}{Using the He\,{\sc ii}\,$\lambda 4686$ line.}
\tablefoottext{e}{Using an average of the He\,{\sc i}\,$\lambda 4472, 5875$ lines.}
\tablefoottext{f}{Using the N\,{\sc iv}\,$\lambda 4060$ line, except for phases $\phi{=}0.54, 0.60$, where the N\,{\sc iii}\,$\lambda \lambda 4634, 4642$ doublet 
is used.}
\tablefoottext{g}{Using an average of the Si\,{\sc iv}\,$\lambda \lambda 4089, 4116$ lines.}

}
\end{center}
\label{tab:RVs}
\end{table}


Several UV spectra were retrieved from the Mikulski Archive for Space Telescopes (MAST).
A flux-calibrated, large-aperture IUE spectrum (swp15908, PI: Burki, $\phi{=}0.18$), covering the range $1150{-}1980\,\AA$ 
with $R{\approx}10\,000$ and S/N${\approx}25$, was used for detailed spectroscopy and fitting of the total spectral energy distribution (SED).
A low dispersion IUE spectrum (lwr04256, PI: Savage, $\phi{=}0.18$) covering $1850{-}3350\,\AA$
with $R{\approx}150$ and S/N${\approx}10$ was used for fitting the SED alone. We used high dispersion Hubble Space Telescope (HST) spectra taken with the High Resolution Spectrograph (HRS)
and the G160M grating 
(Z0NG0402T, Z0Z30202T, Z0Z30402T, Z0Z30602T, PI: Hutchings) that 
cover primarily the C\,{\sc iv}\,$\lambda \lambda 1548, 1551$ doublet ($1528{-}1563\,\AA$, $R{\approx}15\,000$, S/N${\approx}30$), corresponding 
to the phases $\phi=0.99, 0.03, 0.98$, and $0.49$, respectively. Finally, a FUSE spectrum covering $900{-}1190\,\AA$ with $R{\approx}15\,000$ and S/N${\approx}25$ 
was also used for detailed spectroscopy and SED fitting (p1030401000, PI: Sembach, $\phi{=}0.06$).
We checked for possible source contamination of the IUE and FUSE spectra. The IUE large aperture does not cover any sources that can significantly contribute to the 
UV flux. The FUSE spectrum could, in principle, be contaminated by nearby sources, depending on the positioning. However, 
the flux levels of the HST, IUE, and FUSE spectra are consistent within $5\%$, implying that they are not contaminated by other sources.

$UBV$ band photometry was obtained from \citet{Mermilliod2006}. 
Near-infrared photometry ($J, H, K_S$) was obtained from \citet{Cutri2003}, while WISE photometry is available from \citet{Cutri2012}. Finally, IRAC 
photometry was taken from a compilation by \citet{Bonanos2010}.

\section{Identifying the companions of AB\,6}
\label{sec:binquad}

\begin{figure}[]
\centering
  \includegraphics[width=\columnwidth]{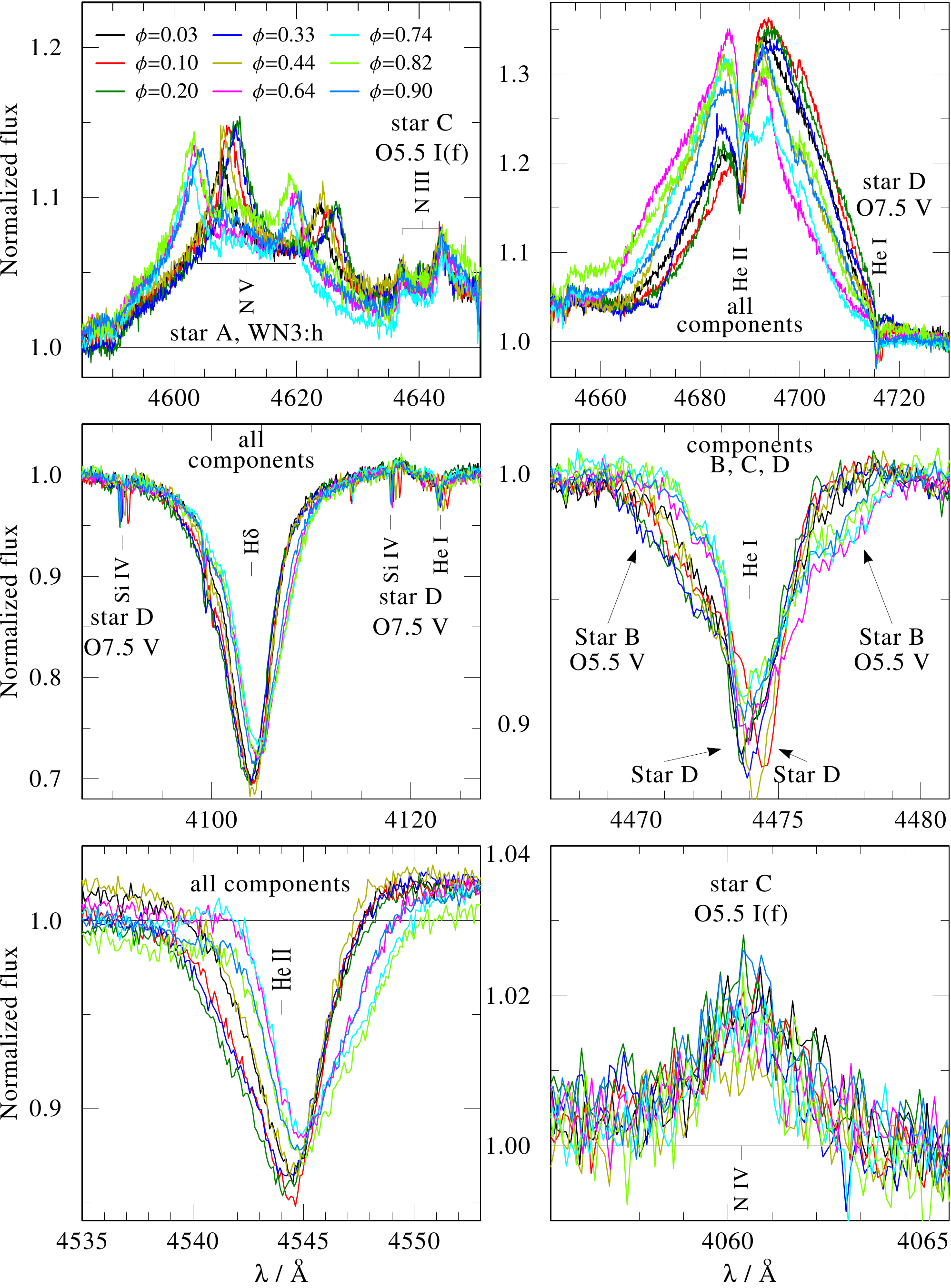}
  \caption{ Zoom-in of the nine available UVES spectra for various spectral lines (see legend). 
  We note the antiphase motion of star B, the static RVs of star C, and the 
  long-term RV variation of star D (colored version available online).
  }
\label{fig:overview}
\end{figure} 

In Fig.\,\ref{fig:overview}, we plot the nine UVES spectra. The spectra are binned at $0.1\,\AA$ for clarity. 
 \citet{Moffat1982} already noted that the relatively high mass ratio they reported ($q{\equiv}M_\text{O}/M_\text{WR}{>}4$) may be a consequence of additional components hidden in the system. 
Our analysis proves that this conjecture was correct (see Fig.\,\ref{fig:quadquint}).

\subsection{Components A (WN3:h) and B (O5.5~V)}
\label{subsubsec:AB}
The WR primary  (WN3:h, see Sect.\,\ref{subsec:specanstarA}) is easy to 
distinguish because of its rapidly moving emission lines (e.g., N\,{\sc v}\,$\lambda \lambda 4604, 4620$ and 
He\,{\sc ii}\,$\lambda 4686$, upper panels of Fig.\,\ref{fig:overview}) , especially seen in the spectra taken close to opposite quadratures 
($\phi{=}0.20,0.74$).  To look for the companion, we ought to identify features that 
portray an antiphase behavior compared to the WR star.
\citet{Foellmi2003SMC} identified such an antiphase motion 
in the Balmer lines owing to their high S/N, 
and used these  lines to derive RVs for the companion. The antiphase motion in H$\delta$ can be seen 
in the {middle} left panel of Fig.\,\ref{fig:overview}. It is evident that the RV amplitude of this line 
is relatively low (${\approx}50\,$\kms), which explains the high mass ratio obtained by \citet{Foellmi2003SMC}. 
However, as it turns out, the Balmer lines contain contributions from four 
stars (Sect.\,\ref{subsubsec:CD}), making them unsuitable for accurate RV measurements.

The true companion 
of the WR star exhibits a much larger RV amplitude than derived by \citet{Foellmi2003SMC}. This can be seen when comparing the  
quadrature spectra in the middle right panel of Fig.\,\ref{fig:overview}, where the He\,{\sc i}\,$\lambda 4471$
line is shown (star B). One can notice that the $\phi{=}0.74$ spectrum exhibits a roundish absorption feature that is  
redshifted with respect to the line center. This same feature, somewhat less pronounced, becomes blueshifted in the $\phi{=}0.20$ spectrum. 
Star B is a rapidly rotating O-type star, which we later classify as O5.5~V (see Sect.\,\ref{subsec:specanstarB}).
Companions A and B together make the 6.5\,d-period WR binary in AB\,6.

\subsection{Components C (O5.5~I(f)) and D (O7.5~V)}
\label{subsubsec:CD}

Figure\,\ref{fig:overview} reveals that 
additional components are present in the spectrum of AB\,6. Star C, which 
contributes to all prominent spectral features, can be  
seen in emission lines that are apparently static (within errors), such as 
N\,{\sc iii}\,$\lambda \lambda 4634, 4642$ and N\,{\sc iv}\,$\lambda 4060$.
These lines belong to neither star A nor star B because they do not follow their Doppler motion.  For the same reason, these lines cannot originate 
in a wind-wind collision (WWC) cone (see Sect.\,\ref{subsec:wwc}).
Hence, star C is an emission line star (O5.5~I(f), see Sect.\,\ref{subsec:specanstarC}) that displays little or 
no Doppler motion.

\begin{figure}[]
\centering
  \includegraphics[width=.95\columnwidth]{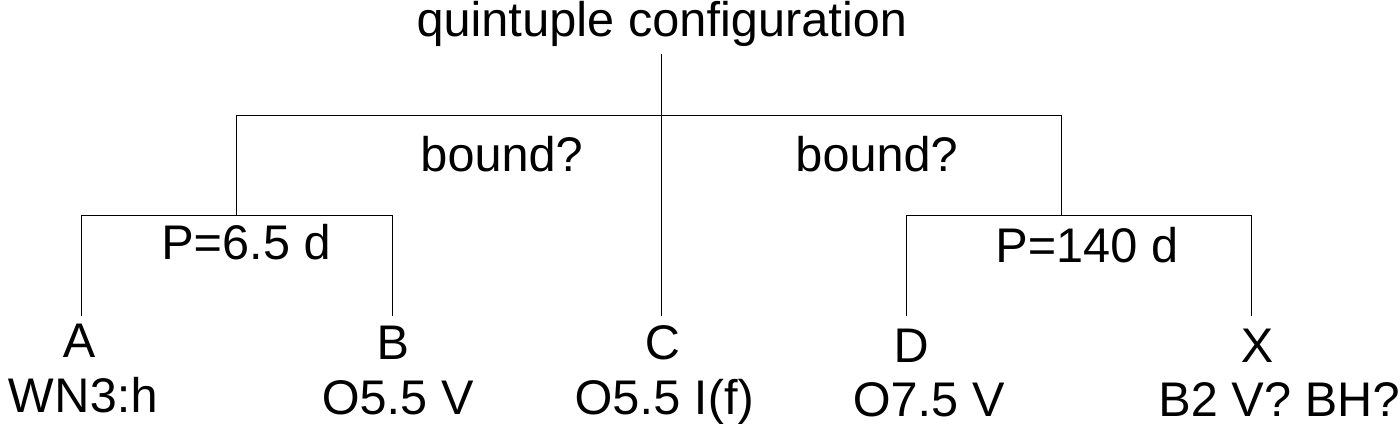}
  \caption{ Most likely configuration of AB\,6 (see Sect.\,\ref{subsec:CD}).
  }
\label{fig:quadquint}
\end{figure}

The fourth component, star D, can be easily seen in the Si\,{\sc iv}\,$\lambda \lambda 4089, 4116$ doublet (middle left panel
in Fig.\,\ref{fig:overview}), but also in  
He lines (mostly He\,{\sc i}). This star, later classified as O7.5~V (Sect.\,\ref{subsec:specanstarD}), exhibits very narrow absorption features and significant RVs. However, 
neither the line profiles nor the RVs  may be attributed to stars A, B, or C. For example, 
it is evident from Fig.\,\ref{fig:overview} that star D remains almost static 
in the two quadrature spectra ($\phi{=}0.74$ and $\phi{=}0.20$), but shows a clear RV shift in another 
spectrum ($\phi{=}0.10$), which was taken ${\approx}80\,$d after the quadrature spectra. A similar behavior 
can be seen in He\,{\sc i} lines and to a lesser extent in the He\,{\sc ii}\,$\lambda 4686$ line. 
In Sect.\,\ref{subsec:CD}, we show that it is very likely that star D forms a second binary system with a fifth component that is not seen in our spectra, 
making AB\,6 a quintuple system (Fig.\,\ref{fig:quadquint}). 

\section{Orbital analysis}
\label{sec:orban}

\subsection{RV measurements}
\label{subsec:RVmes}

To measure the RVs of the various components in the nine available UVES spectra, 
we employed a technique identical to that described in \citet{Shenar2017a}. The method relies on cross-correlating 
the spectra with a template spectrum. 
If the line that is to be cross-correlated originates primarily in one component,
the best template  can be constructed 
by co-adding the observations in the frame of reference of this companion. However, to do so, one needs a 
first estimate for the RVs. This first estimate was obtained using  preliminary model spectra
(see Sect.\,\ref{sec:specan}). 


For star A (WN3:h),  we measured RVs for both the He\,{\sc ii}\,$\lambda 4686$ and 
N\,{\sc v}\,$\lambda \lambda 4604, 4620$ lines. The latter clearly offers  a more reliable way of measuring 
the RVs. First, the sharply peaked N\,{\sc v} profiles allow for a relatively accurate RV determination. Second, the N\,{\sc V} lines 
form close to the stellar surface ($r\approx 1.5\,R_*$) and their Doppler shift should therefore represent the motion of the 
WR star much more accurately than the He\,{\sc ii}\,$\lambda 4686$ line. 
Third, the N\,{\sc v} doublet is not contaminated by the other stars. We also measured the RVs of the N\,{\sc v}\,$\lambda 4944$ line. The values agree with 
the N\,{\sc v} doublet within errors (see Fig.\,\ref{fig:phasefit}). Because of the relatively low S/N of the N\,{\sc v}\,$\lambda 4944$ line and normalization 
uncertainties, we adopted the N\,{\sc v}\,$\lambda \lambda 4604, 4620$  RVs for our orbital solution (see Sect.\,\ref{subsec:orbit}).

\begin{figure}[]
\centering
  \includegraphics[width=\columnwidth]{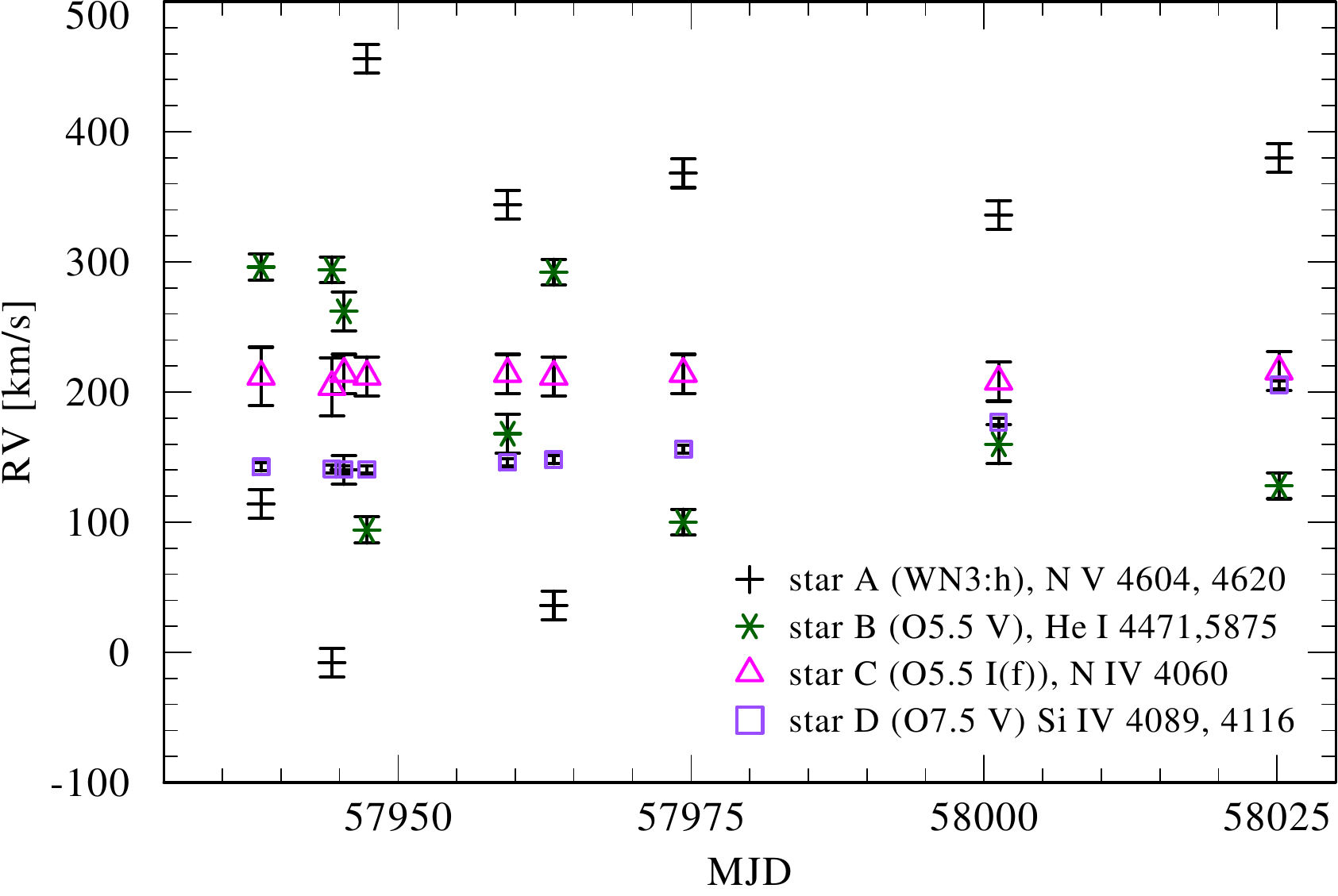}
  \caption{Derived RVs for the components of AB\,6 (see legend). 
}
\label{fig:RVs_all}
\end{figure} 

For star C (O5.5~I(f)), we used the isolated N\,{\sc iv}\,$\lambda 4060$ line, since it is not contaminated by  
the other stars and is relatively easy to cross-correlate with. Virtually identical RVs were obtained using the N\,{\sc iii} lines.
For star D (O7.5~V), we used an average of the narrow Si\,{\sc iv}\,$\lambda \lambda 4089, 4116$ lines.
The same method cannot be easily applied to star B, since all its spectral features are contaminated by the other components.
However, we took advantage of the fact that the RVs of the other components are known. For each of our 
nine spectra, we subtracted preliminary models for stars A, C, and D to isolate the contribution of star B. We then used the He\,{\sc i} 
$\lambda 4471$ and $\lambda 5875$ lines for the cross-correlation, which have the advantage of not being contaminated by star A.

Once these estimates were obtained, we constructed four new templates by co-adding the observations in the  frame
of stars A, C, and D, respectively. These templates were used exactly as described above. The RVs obtained in this way  
were found to be in good agreement with the previous iteration, but the scatter was smaller. The derived RVs for the nine spectra 
and the four components are shown in Fig.\,\ref{fig:RVs_all} and are compiled in Table\,\ref{tab:RVs}.

\subsection{Orbital solution for the WR binary (stars A+B)}
\label{subsec:orbit}

Using the RVs measured for stars A and B, we can now derive an orbital solution for the system. 
For the WR component, we also make use of RVs measured by \citet{Moffat1982}, 
\citet{Hutchings1984}, and \citet{Foellmi2003SMC} for the He\,{\sc ii}\,$4686$ line. Unfortunately, \citet{Foellmi2003SMC} did not tabulate their measured RVs, so 
we extracted the RVs from figure 3 in their study using the {\it Engauge Digitizer Software}\footnote{Mark Mitchell, Baurzhan Muftakhidinov, and Tobias Winchen et al., 
markummitchell.github.io/engauge-digitizer}, accounting for the systematic instrumental 
shifts reported in their work. 
With the full set of RVs, we derive 
an orbital solution using a self-written Python script, which uses the standard Python routine lmfit\footnote{lmfit.github.io/lmfit-py/}.
The routine fits both components simultaneously for the period $P$, time of inferior WR conjunction $E_0$ (i.e., the WR star is in front the O star), 
RV amplitudes $K_\text{WR}$ and $K_\text{O}$, eccentricity 
$e$, argument of periastron $\omega$, and systematic velocity $V_0$. 
Since deriving absolute RV values using emission lines is prone to significant systematic errors originating in the uncertain
velocity law and mass-loss rate, we allow for a constant shift parameter for the RVs derived for the WR star. The 
shift parameter of the UVES measurements obtained from the final fit is given in the footnotes of Table\,\ref{tab:RVs}

One concern is that the RVs from previous studies were measured using the He\,{\sc ii}\,$\lambda 4686$
line. Since this line forms a few stellar radii above the stellar surface, it is more susceptible to 
distortions (e.g., WWCs, see Sect.\,\ref{subsec:wwc}) and is generally known to exhibit a strong variability that is both phase and epoch 
dependent \citep[e.g.,][]{Foellmi2008, Koenigsberger2010}. Therefore, we fit two orbital solutions. For both solutions, 
He\,{\sc i}\,$\lambda 4471, 5875$ are used for star B, but different lines are used for star A.

\begin{figure*}[]
\centering
  \includegraphics[width=\textwidth]{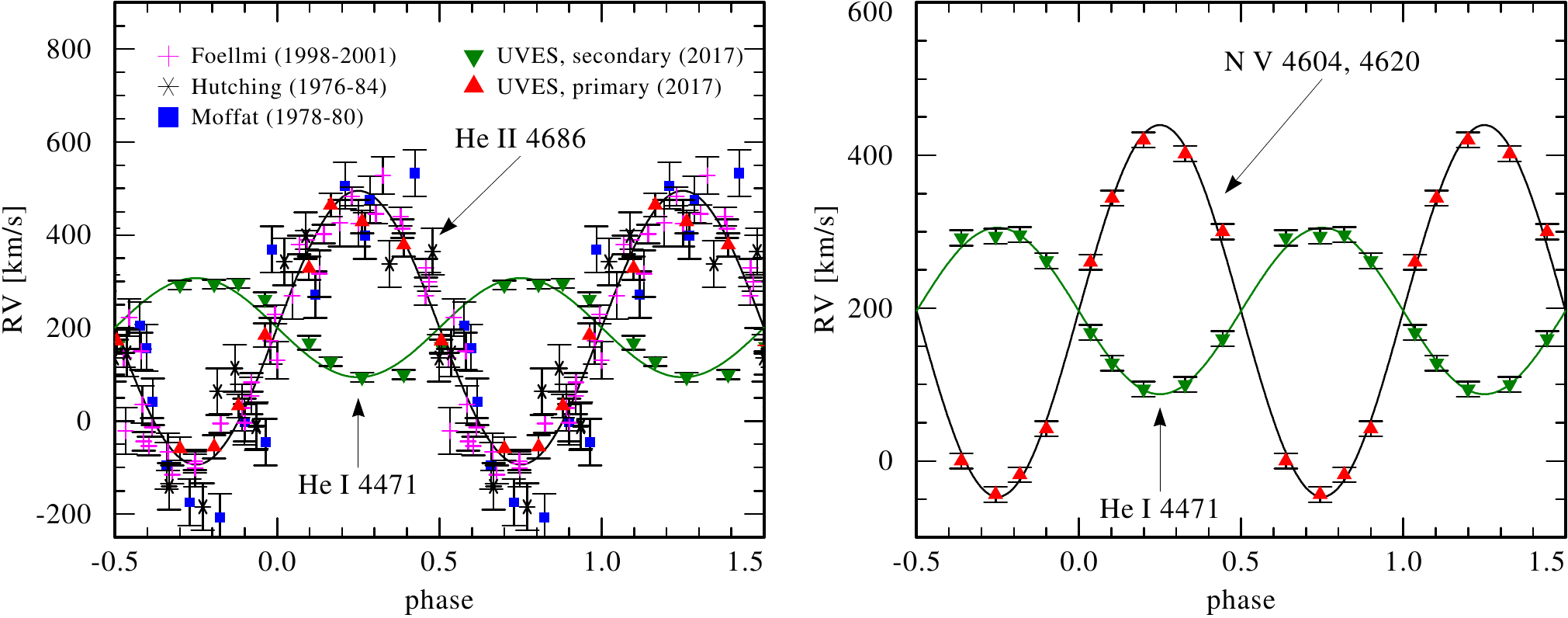}
  \caption{\emph{Left panel:} SB2 orbital solution for the WR binary in AB\,6 (WR: black curve; O: green curve) using 
  the complete set of RV measurements of the He\,{\sc ii}\,$\lambda 4686$ line for star A and the He\,{\sc i}\,$\lambda 4471$ line 
  for star B (see legend). \emph{Right panel:} Same as left panel, but using the N\,{\sc v}\,$\lambda \lambda 4604, 4620$ doublet 
  RVs for star A. See text for details
}
\label{fig:orbitfit}
\end{figure*}

For the first solution, we use the He\,{\sc ii}\,$\lambda 4686$ line, combining 
old${+}$new RVs. The solution ($\chi^2{=}3.8\,${\kms}) is shown on the left panel of Fig.\,\ref{fig:orbitfit}. 
A good agreement is obtained between the old and new measurements and enables 
an accurate derivation of the period, which is found to be $P = 6.53840\,$d. However, it is also clear that a phase shift exists between 
the RVs of stars A and B, implying that the He\,{\sc ii}\,$\lambda 4686$ line does not represent the true motion of the WR star. This is most likely a consequence 
of WWC  (see Sect.\,\ref{subsec:wwc}).

As discussed above, the N\,{\sc v} RVs should provide a better representation of the true motion of star A. For the second solution, we therefore 
fit the nine measured RVs of the N\,{\sc v} doublet for star A simultaneously with  star B. Because the 
period can be much better constrained using the He\,{\sc ii}\,$\lambda 4686$ orbital solution, which covers ${\approx}40$ years of data, we fix it to the 
value given above. We then use the measured N\,{\sc v} RVs to fit for the remaining orbital parameters.
The orbital fit is shown on the right panel of Fig.\,\ref{fig:orbitfit} and has a $\chi^2{=}0.6\,${\kms}, which is six times better than 
obtained for the He\,{\sc ii}\,$\lambda 4686$ fit. This time, no phase shift is seen between the two components, suggesting that this solution represents the 
true orbital configuration of the system much better. Therefore, we adopt the orbital parameters obtained using the 
N\,{\sc v} doublet. We note that the two solutions result in similar parameters, except for $K_\text{WR}$, which is about 30\,{\kms} larger in the He\,{\sc ii}
solution.
The final parameters are given 
in Table\,\ref{tab:orbpar}, along with values obtained by \citet{Foellmi2003SMC} for comparison. 

\begin{table}[]
\caption{Derived orbital parameters for the WR binary A+B}
\begin{center}
\begin{tabular}{l   c  c}
\hline                                                                                              
Parameter                         & Our results\tablefootmark{a}         & \citet{Foellmi2003SMC}       \\ 
\hline                                                                                     
$P_\text{orb}$ [days]             & $6.53840{\pm}0.00004$ & $6.5364{\pm}0.0007$      \\
$E_0$ [MJD]                       & $51924.17\pm0.08$\tablefootmark{b}  & $51920.9\pm0.2$\tablefootmark{c}                         \\
$K_\text{WR}$ [\kms]              &  $243\pm 4$                    &  $290\pm 10$                  \\ 
$K_\text{O}$ [\kms]               &  $109\pm 4$                    &  $66\pm10$                    \\
$e$                               &  $0$\tablefootmark{d} (fixed)  &  $0.10\pm 0.03$               \\
$\omega [^\circ]$                 & n/a                           &  $103\pm 20$                  \\
$M_\text{WR} \sin^3 i$ [$M_\odot$] & $9.2\pm0.5$                   & $5.6\pm 0.4$                   \\
$M_\text{O} \sin^3 i$ [$M_\odot$] & $20.4\pm 0.9$                  & $24.5\pm 1.6$                  \\
$q{=}M_\text{O} / M_\text{WR}$    & $2.2\pm 0.1$                  & $4.4\pm 0.4$                      \\
$a_\text{WR} \sin i$ [$R_\odot$]          & $31.5\pm0.5$                     & $37.5\pm1.3$                   \\
$a_\text{O} \sin i$ [$R_\odot$]          & $14.1\pm 1.1$                    & $8.5\pm 1.3$                    \\
$V_0$ [\kms]                      & $196\pm4$                      & $199\pm3$                      \\                                           
$i$ [$^\circ$]                    & $53^{+8}_{-5}$\,\,\tablefootmark{e}  & $65\pm7$\tablefootmark{f}    \\                            
$M_\text{orb, WR}$ [$M_\odot$]     & $18^{+5}_{-5}$                     & $7.5\pm1.5$                 \\
$M_\text{orb, O}$ [$M_\odot$]     & $41^{+10}_{-10}$ (adopted)\tablefootmark{g}  & $33\pm6$                    \\
$a_\text{WR} $ [$R_\odot$]                & $40\pm10$                        & $43$                              \\
$a_\text{O} $ [$R_\odot$]                & $18\pm5$                        & $10$                             \\
$R_\text{RL, WR}$ [$R_\odot$]    & $18\pm3$                        & $13\pm2$                              \\
$R_\text{RL, O}$ [$R_\odot$]    & $26\pm5$                       & $26\pm10$                             \\
\hline
\end{tabular}
\tablefoot{
\tablefoottext{a}{$P$ obtained from the He\,{\sc ii}\,$\lambda 4686$ fit (left panel of Fig.\,\ref{fig:orbitfit}). The remaining 
parameters are obtained from the N\,{\sc v}  fit  (right panel of Fig.\,\ref{fig:orbitfit})}
\tablefoottext{b}{Defined so that $\phi{=}0$ is at inferior WR conjunction (WR star in front of O star).}
\tablefoottext{c}{Time of periastron passage.}
\tablefoottext{d}{Formal fit yielded $e=0.03\pm0.02$.}
\tablefoottext{e}{Implied from $M_2 \sin^3 i$ by adopting $M_2{=}41{\pm}10\,M_\odot$}
\tablefoottext{f}{From unpublished polarimetric data.}
\tablefoottext{g}{From calibration with evolutionary masses (see text)}
}
\end{center}
\label{tab:orbpar}
\end{table}

The period derived in this work agrees with previous derivations within 3$\sigma$. Since our derived value is based on data extending 
over 40 years, it should be closer to the true period. The eccentricity is found to be smaller ($0.03\pm0.02$) than previously reported, implying 
a virtually circular orbit.  
The RV amplitude of star B, $K_\text{O}$, is found to be almost twice as large as previously reported. As a consequence, the mass ratio $q{=}M_\text{O}/M_\text{WR}$ diminishes 
from 4.4 to 2.2. 
Since only $M\sin^3 i$ can be derived from an orbital analysis, 
the orbital masses strongly depend on $i$. 
\citet{Foellmi2003SMC} reported 
$i{=}65\pm7^\circ$ based on unpublished polarimetric data, but as this analysis was never published, it cannot 
be readily adopted. 

In Sect.\,\ref{subsec:lc}, we constrain $i$ using the light curve of the system. However, our light-curve model is too crude 
to yield significant constraints on the orbital masses. Instead, given that the parameters of star B are fairly well 
known, we adopt its mass from calibration with evolution models. 
For this purpose, we use the BONNSAI\footnote{The BONNSAI web-service is available at www.astro.uni-bonn.de/stars/bonnsai} Bayesian statistics tool \citep{Schneider2014}. 
The tool interpolates over evolutionary tracks calculated at SMC metallicity by \citet{Brott2011}
for stars with initial masses up to $100\,M_\odot$ and over a wide range of initial rotation velocities. Using the derived values for $T_*$, $\log L$, 
and $\log g$, and their corresponding errors, BONNSAI predicts $M_\text{O}{=}41{\pm5}\,M_\odot$. This is $4\,M_\odot$ lower than the evolutionary mass 
of a star with the same parameters at solar metallicity, which is consistent with theoretical expectation \citep[e.g.,][]{Markova2009, Garcia2013}. This value agrees with the spectroscopic 
mass of star B, $M_\text{spec}$, within errors (which are very large). Considering the 
uncertainties involved in this approach, 
we conservatively adopt $M_2{=}41{\pm}10\,M_\odot$ for star B. This implies $i{=}53^\circ$ and $M_\text{WR}{=}18\,M_\odot$.
In Table\,\ref{tab:orbpar}, we also give the 
semimajor axes $a$ and the Roche lobe radii $R_\text{RL}$, calculated 
using the Eggleton approximation \citep{Eggleton1983}.



\subsection{Light curve modeling}
\label{subsec:lc}

%

\begin{figure}[]
\centering
  \includegraphics[width=\columnwidth]{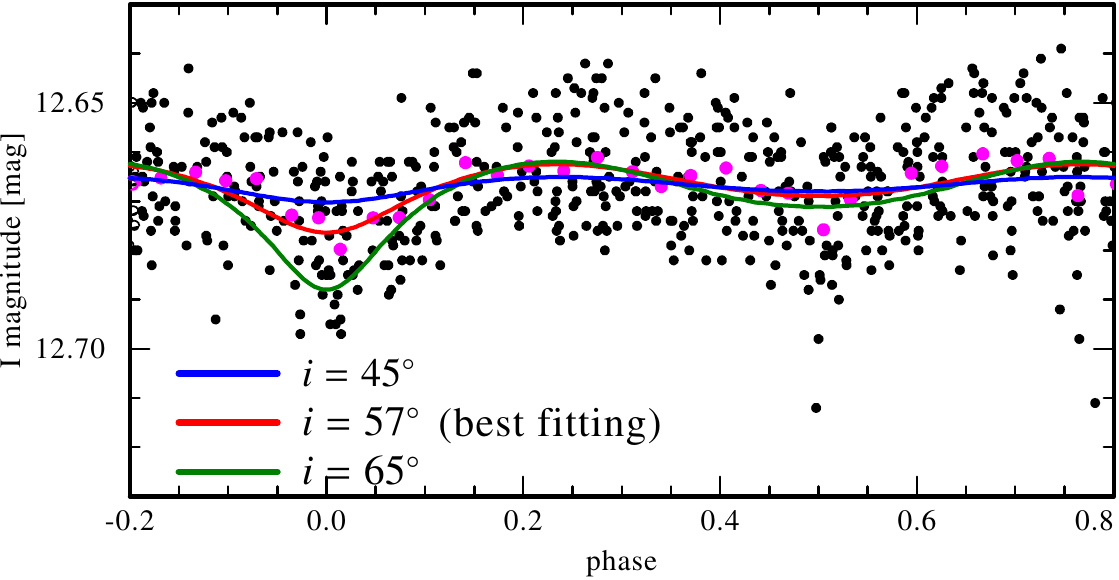}
  \caption{Combined Phoebe + wind-eclipse light curve model calculated for three inclinations ($i{=}45^\circ, 65^\circ,$ and 
  the best-fitting value $57^\circ$; see legend), compared to the OGLE light curve (unbinned and binned at $\Delta \phi{=}0.03$). Formal errors 
  in the unbinned light curve are comparable to the symbol sizes
}
\label{fig:lc-fit}
\end{figure}

In Fig.\,\ref{fig:lc-fit}, we show the $I$-band  OGLE light curve 
of AB\,6 \citep{Udalski1998}, phased with the ephemeris in Table\,\ref{tab:orbpar}. 
Two faint and broad dips can be seen during both conjunctions ($\phi{=}0,0.5$). 
In principle, these could be grazing photospheric eclipses, wind eclipses, 
ellipsoidal variations, reflection effects, or even  
induced pulsations \citep[e.g.,][]{Pablo2015}. 

At first, it seems that the two dips are suggestive of grazing eclipses. However, given the large temperature difference between stars A and B, the eclipse of 
the WR star ($\phi{=}0.5$) should be significantly more pronounced than that at $\phi{=}0$, which is not the case. 
Moreover, grazing eclipses should be relatively narrow, with $\Delta \phi 
{\approx} 0.05{-}0.1$ \citep[e.g., \object{V444 Cyg},][]{Antokhin1995}, while the dips observed here spread over $\Delta \phi{\approx}0.3$. Lastly, given the orbital parameters and 
the radii derived from our spectral analysis (Sect.\,\ref{sec:specan}),
grazing eclipses are only obtained for $i > 73^\circ$, resulting in implausible masses of $M_\text{O}{<}23\,M_\odot$ and $M_\text{WR}{<}10\,M_\odot$. 
It therefore seems likely that other mechanisms are at work here.

A natural explanation for the dip at $\phi{=}0$ (WR star in front) are wind eclipses \citep{Lamontagne1996, St-Louis2005}. The relatively strong wind 
of the WR star can scatter the light from its companion, primarily via electron scattering. This can be modeled in a straight forward fashion by assuming 
a simple $\beta$ law for the velocity field of the WR star and adopting the parameters derived in this study, including the mass-loss rate, which is derived in 
Sect.\,\ref{sec:specan}.
The formalism has been thoroughly described in \citet{Lamontagne1996} and  \citet{Munoz2017}. 

Yet wind eclipses cannot explain the dip observed around $\phi{=}0.5$, as the wind of star B is too weak to lead to a significant eclipse. To investigate possible mechanisms
that can reproduce this dip, we generated a light curve model using the Physics Of Eclipsing BinariEs 1.0 (PHOEBE) program \citep{Prsa2005}. While this
program is robust, it does not include wind physics. However, it is still useful for determining plausible physical models that could influence the light curve. 
Using the parameters derived in Tables\,\ref{tab:orbpar} and \ref{tab:specan}, our model light curves suggest that the dominant modulation 
is heating of the O star surface by the WR star, referred to as the
reflection effect \citep{Kopal1954}. 
When this heated up surface points toward the observer ($\phi{=}0$), excess emission is predicted, while an emission deficiency is predicted at $\phi{=}0.5$ - exactly as observed.
While this phenomenon is likely more complex in WR stars than our model is able to replicate,
we are able to recreate both the shape and magnitude of the observed variation simply by adopting the parameters of the system, which lends credence
to our rough approximations.

By interpolating over reflection models calculated with PHOEBE at various inclinations and combining this with the wind-eclipse model described above,
we generate a combined Phoebe+wind-eclipse light curve model, which can be calculated at a given inclination $i$. All the relevant parameters except for the inclination 
are kept fixed, and the constant contribution of stars C and D is accounted for in the model. The best-fitting model, obtained for 
an inclination of $i{=}57^\circ$, is shown in Fig.\,\ref{fig:lc-fit}.
We also show the models for 
$i{=}45^\circ$ and $i{=}65^\circ$ for comparison. Since the wind eclipse at $i{=}65^\circ$ overestimates the absorption at $\phi{\approx}0$, we can rule out grazing 
eclipses in the system, as such eclipses would yield additional absorption to the wind eclipse.
The formal error on $i$ is ${\pm}2^\circ$, but given all the assumptions involved in this model, this error is clearly underestimated. 
This result should be considered as a qualitative illustration of how reflection + wind eclipses can reproduce the main features of the light curve at an inclination that is 
consistent with the adopted value of $i{=}53^\circ$.

\subsection{Wind-wind collisions}
\label{subsec:wwc}

When two stars in a binary possess strong stellar winds, the winds are expected to collide and 
form a shock cone around the companion whose mass loss is smaller \citep{Stevens1992, Moffat1998}. The shocked material 
radiates in one or more spectral bands (X-rays, visual) as it streams along the cone, which corotates with the system. 
Some known examples for WWC 
systems are \object{$\eta$ Car} \citep{Parkin2009} and $\gamma^2$ Vel \citep{Richardson2017} in the Galaxy, 
\object{HD\,5980}  in the SMC \citep{Naze2007}, and \object{BAT99 119} in the LMC \citep{Shenar2017a}. Unless the system is seen close to pole-on, and neglecting aberration and 
Coriolis forces, 
the emission from the WWC cone should become redshifted at $\phi{=}0$ (the stream recedes away from the observer)
and blueshifted at $\phi{=}0.5$  \citep[e.g., figure 6 in][]{Hill2000}. 

\begin{figure}[]
\centering
  \includegraphics[width=\columnwidth]{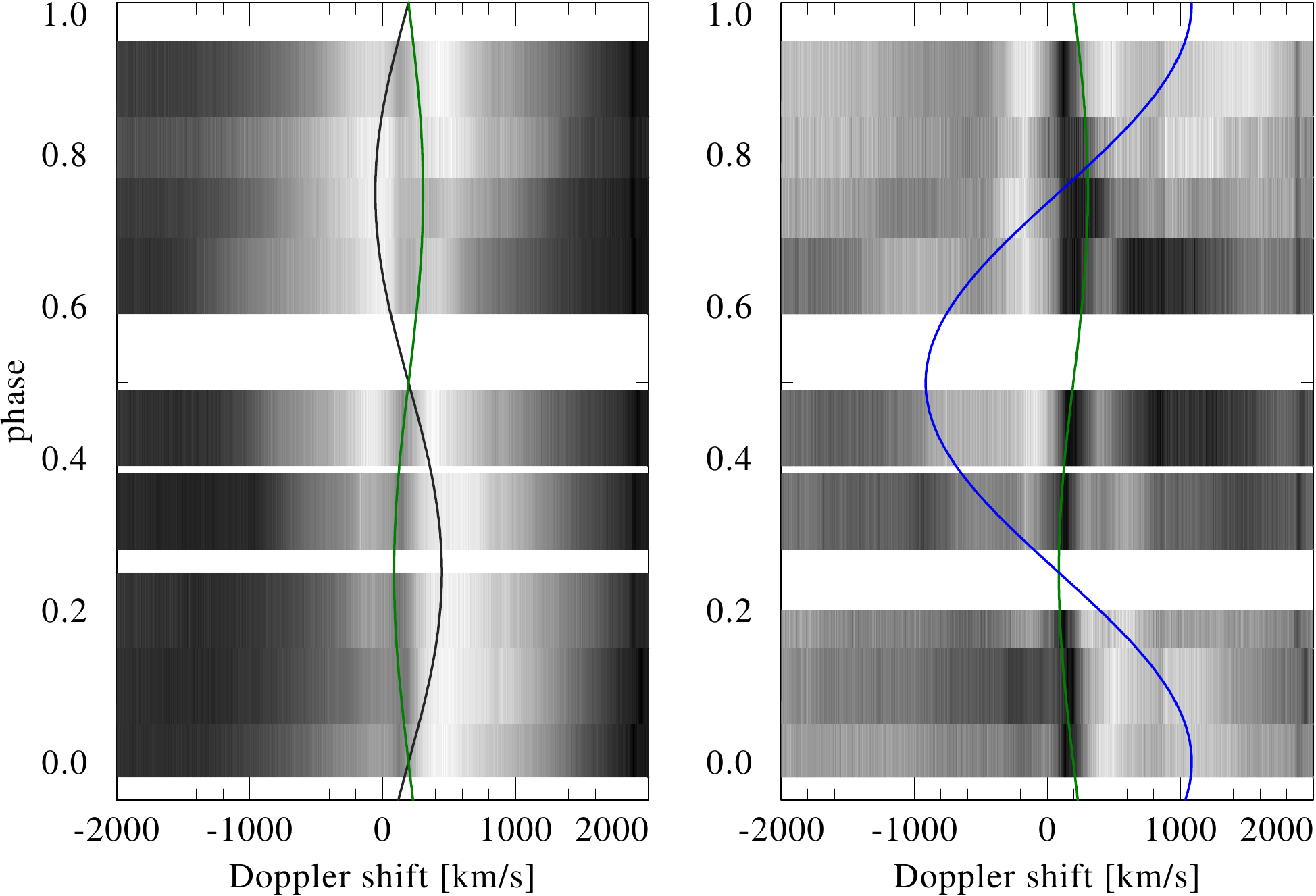}
  \caption{\emph{Left panel:} Dynamical spectrum of the He\,{\sc ii}\,$\lambda 4686$ line in velocity space. The RV curves of stars A and B  are 
  shown. \emph{Right panel:} Same as left panel, but with the contribution of the WR star subtracted. The residual emission excess (traced with 
  the blue curve) follows the expected behavior of a WWC cone.
}
\label{fig:wwc}
\end{figure} 

Evidence for WWC can be seen in Fig.\,\ref{fig:wwc}, where a dynamical spectrum of the He\,{\sc ii}\,$\lambda 4686$ line using the nine UVES observations is shown. We show both the original 
normalized spectra (left panel) and spectra from which the average contribution of the WR star,  shifted to its respective RV, was subtracted (right panel). The average was constructed by co-adding
the observations in the frame of reference of the WR star. 

After subtracting the contribution of the WR star, a clear emission excess pattern is visible that is redshifted at $\phi{=}0$ and 
blueshifted at $\phi{=}0.5$ (traced with a blue sine curve), 
exactly as expected in the case of WWC systems. With significant phase coverage, one can constrain kinematic information about the shock cone from such patterns 
\citep{Luehrs1997, Shenar2017a}. However, with only nine observations and considering the quintuple nature of the system, 
this would not yield any meaningful constraints, and we therefore refrain from such an analysis.

\section{Spectral analysis}
\label{sec:specan}

\begin{figure}[]
\centering
  \includegraphics[width=0.5\textwidth]{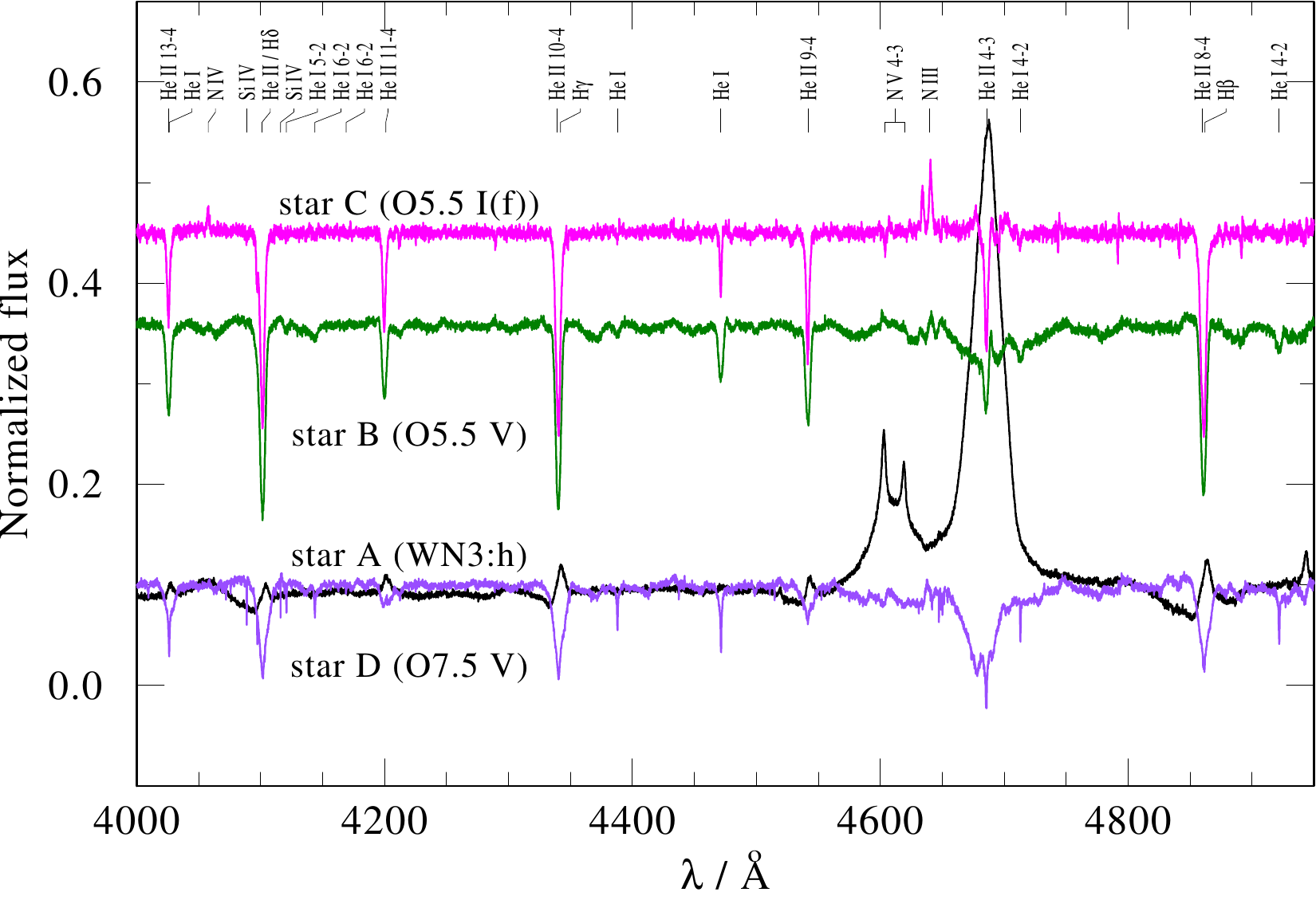}
  \caption{Disentangled spectra of stars A, B, C, and D, shifted to their relative contribution in the visual. 
  With only nine observations and given the four stellar components, 
  the reproduced strengths of the Balmer lines and important He lines are poorly constrained and should not be considered as real.}
\label{fig:disfit}
\end{figure}

Phase-dependent spectra can be used to disentangle them from the constituent spectra \citep{Hadrava1995, Shenar2017a}. 
We attempted to disentangle the UVES spectra using the shift-and-add technique \citep{Marchenko1998b}, which we extended for the case of 
four components. 
The results are shown in Fig.\,\ref{fig:disfit}.
While the disentangled spectra are plausible for most spectral lines, the Balmer 
lines and strong He\,{\sc ii} lines (especially $\lambda 4686$) are poorly constrained. 
We therefore do not rely on these disentangled spectra for the spectral analysis.  However, the disentangled spectra enable us to spectroscopically classify the components.
We use classification schemes by \citet{Smith1996} for star A,  
and quantitative classification schemes by Sana et al.\ (in prep.), which are 
extensions of schemes by \citet{Mathys1988, Mathys1989}, \citet{Walborn1990}, and \citet{Walborn2002} for 
stars B and D. Star C is classified morphologically using \citet{Massey2009}. Details are given in Sects.\,\ref{subsec:specanstarD}-\ref{subsec:specanstarA}.
Thankfully, important spectral features can be unambiguously attributed to each component
(Fig.\,\ref{fig:overview}). The spectral analysis thus virtually reduces to the analysis of four single stars 
with the exception of the unknown light ratios.

The spectral analysis is performed with the Potsdam Wolf-Rayet\footnote{{\sc PoWR} models of WR and OB-type stars can be downloaded at
www.astro.physik.uni-potsdam.de/PoWR, \citep{Todt2015}} ({\sc PoWR}) model atmosphere code \citep{Graefener2002, Hamann2004}, 
which is applicable to any hot star \citep[e.g.,][]{Gvaramadze2014, Gimenez2016}. 
The code iteratively solves the comoving frame, nonlocal thermodynamic equilibrium (non-LTE) radiative transfer 
and statistical balance equations in spherical symmetry under the constraint of energy conservation, yielding
the population numbers in the photosphere and wind. By comparing output synthetic spectra to observed spectra, fundamental stellar parameters are derived.
A detailed description of the code  is given by \citet{Graefener2002} and \citet{Hamann2004}. Only essentials are given here.

Aside from the chemical abundances and the wind velocity field, 
a PoWR model is defined by four fundamental stellar parameters: the effective temperature $T_*$, 
stellar luminosity $L$, mass-loss rate $\dot{M}$, and surface gravity $g_*$, the latter being seldom important for WR stars. 
The effective temperature is defined at the stellar radius $R_*$ via the Stefan-Boltzmann equation $L{=}4\,\pi\,\sigma\,R_*^2\,T_*^4$. 
The stellar radius is defined at the inner boundary of the model, fixed at Rosseland mean optical depth of 
$\tau_\text{Ross}{=}20$. The outer boundary is set to $R_\text{max}{=}1000\,R_*$.
The gravity $g_*$ relates to the radius $R_*$ and mass $M_*$  via the usual definition, $g_*{=}G\,M_*\,R_*^{-2}$. 

\renewcommand{\arraystretch}{1.1}
\setlength{\tabcolsep}{1.3mm}

\begin{table}[]
\scriptsize
\small
\caption{Derived parameters from the spectral analysis of AB\,6}
\label{tab:specan}
\begin{center}
\begin{tabular}{l | c  c c c}
\hline      
Parameter & star A & star B & star C & star D  \\ 
\hline
Spectral type                                  & WN3:h                   & O5.5~V                            & O5.5~I(f)             & O7.5~V                                                      \\
$T_*$ [kK]                                     & $80^{+20}_{-5}$     & $41.5{\pm}1.0$                    & $37{\pm}2$    & $33{\pm}1$                                     \\
$\log g$ \gcgs                                  & -                      & $4.0{\pm}0.3$                     & $3.6{\pm}0.2$     & $4.0{\pm}0.3$                                      \\
$\log L$ [$L_\odot$]                           & $5.87{\pm}0.15$        & $5.65{\pm}0.10$                    & $5.75{\pm}0.10$   &  $4.88{\pm}0.05$                                            \\ 
$\log R_\text{t}$ [$R_\odot$]                  & $1.1^{+0.1}_{-0.4}$  &  -                              & -                   &  -                                               \\
$M_\text{spec}$\,$[M_\odot]$                   & -                      & $61_{-30}^{+60}$               & $47_{-25}^{+50}$    & $26_{-13}^{+41}$                                      \\
$v_\infty$ [\kms]                              & $2000{\pm}100$         & $2000{\pm}300$                    & $2000{\pm}200$       & $2000$                                           \\
$R_*$ [$R_\odot$]                              & $4.7^{+1.5}_{-2.3}$  & $13{\pm}2$                   & $18^{+5}_{-3}$    & $8.4{\pm}1.0$                                       \\ 
$\log \dot{M}$ [$M_\odot\,{\rm yr}^{-1}$]      & $-5.2{\pm}0.2$       & $-6.8{\pm}0.5$                    & $-6.3{\pm}0.3$       & $-9.4$                                         \\
$v_\text{eq} \sin i$ [\kms]                    & $< 100$                 & $210{\pm}10$                      & $90{\pm}20$          & $\le 3$                                                       \\
$v_\text{mac}$ [\kms]                           & -                     & $20$                            & $40$                     & $\le 3$                                                       \\
$\xi_\text{ph}$ [\kms]                         & 100                   & 20                                  & $20$                 & $\le 3$                                                       \\
$v_\text{eq}$ [\kms]                           & <130                  & $265{\pm}30$\tablefootmark{a}        &  -                 &  -                     \\
$X_\text{H}$ (m.f.\tablefootmark{b})           & $0.25{\pm}0.05$         & 0.73                             & 0.73               & 0.73                                           \\
$X_\text{C} / 10^{-4}$ (m.f.)                 & $0.3{\pm}0.2$          & 2.1                               & $0.2\pm0.1$       & $2.1$                                         \\
$X_\text{N} / 10^{-4}$ (m.f.)                 & $80{\pm}20$       & $0.33$                            & $20\pm10$   & $0.33$                                        \\
$X_\text{O} / 10^{-4}$ (m.f.)                   & $\lesssim 0.1$        & $11.3$                           & $11.3$             & $11.3$                                   \\ 
$M_\text{V,John}$[mag]                          & $-4.3{\pm}0.4$        & $-5.8{\pm}0.2$                    & $-6.0{\pm}0.2$     & $-4.36{\pm}0.10$                                  \\
$f_\text{V} / f_\text{V,tot}$                   & 0.09                  & 0.36                               & 0.45               & 0.10                                \\
$\log Q_\text{H}\,[\text{s}^{-1}]$              & $49.8{\pm}0.2$          & $49.1{\pm}0.1$                  & $49.4{\pm}0.1$            & $48.2{\pm}0.1$                                  \\
$E_{B-V}$ [mag]                                 & \multicolumn{4}{c}{$0.08{\pm}0.01$}                                                                                    \\
\hline
\end{tabular}
\tablefoot{
Values without errors were adopted. 
\tablefoottext{a}{Calculated assuming $i$ is aligned with the rotation axis.}
\tablefoottext{b}{m.f.\ = mass fraction}
}
\end{center}
\end{table}

The PoWR models include complex model atoms of H, He, C, N, O, Mg, Si, P, S,  and the iron group elements dominated 
by Fe. Abundances that cannot be derived from the spectra are fixed based on studies by 
\citet{Korn2000}, \citet{Trundle2007}, and \citet{Hunter2007}, which are (in mass fractions)
$X_\text{H}{=}0.73$, $X_\text{C}{=}2.1\cdot10^{-4}$, $X_\text{N}{=}3.26\cdot10^{-5}$, 
$X_\text{O}{=}1.13\cdot10^{-3}$, $X_\text{Mg}{=}9.9\cdot10^{-5}$, $X_\text{Si}{=}1.3\cdot10^{-4}$, 
and $X_\text{Fe} {=}3\cdot10^{-4}$. The remaining elements are scaled to 1/7 solar:
$X_\text{P}{=}8.3\cdot 10^{-7}$ and $X_\text{S}{=}4.4\cdot10^{-5}$.

Quasi-hydrostatic equilibrium is assumed in the subsonic 
regime \citep{Sander2015}, while a $\beta$ law \citep{CAK1975},

\begin{equation}
 v(r) = v_\infty \left(1 - \frac{r_0}{r} \right)^\beta,
 \label{eq:betalaw}
\end{equation}
is assumed for the supersonic regime. Here, $\beta$ is a constant on the order of unity, $v_\infty$ is the terminal velocity, and $r_0$ is a constant chosen so that a smooth transition between the sub- and supersonic regimes is obtained. We adopt 
$\beta{=}0.8$ for OB-type models \citep{Kudritzki1989}. For the WR-star model, we adopt a two-component $\beta$ law (see Sect.\,\ref{subsec:specanstarA})

During the main iteration of the PoWR code, the opacity and emissivity profiles are calculated with a constant 
Doppler width of $v_\text{Dop}{=}30\,$\kms\,for OB-type stars and $100\,$\kms\,for the WR star. In the 
calculation of the emergent spectrum in the observer's frame, 
$v_\text{Dop}$ is calculated from the microturbulence and thermal motion via $v_\text{Dop}{=}\sqrt{\xi^2(r) + v_\text{th}^2(r)}$. The 
microturblence grows from its photospheric value $\xi_\text{ph}$ in proportion to the wind velocity as $\xi(r){=}0.1\,v(r)$. The value $\xi_\text{ph}$ is set to 20\,\kms\, for 
stars B and C, 100\,\kms\, for star A,  and $\xi_\text{ph}{=}3\,${\kms} for star D (see Sect.\,\ref{subsec:specanstarD}).

\begin{figure*}
\centering
  \includegraphics[width=.91\textwidth]{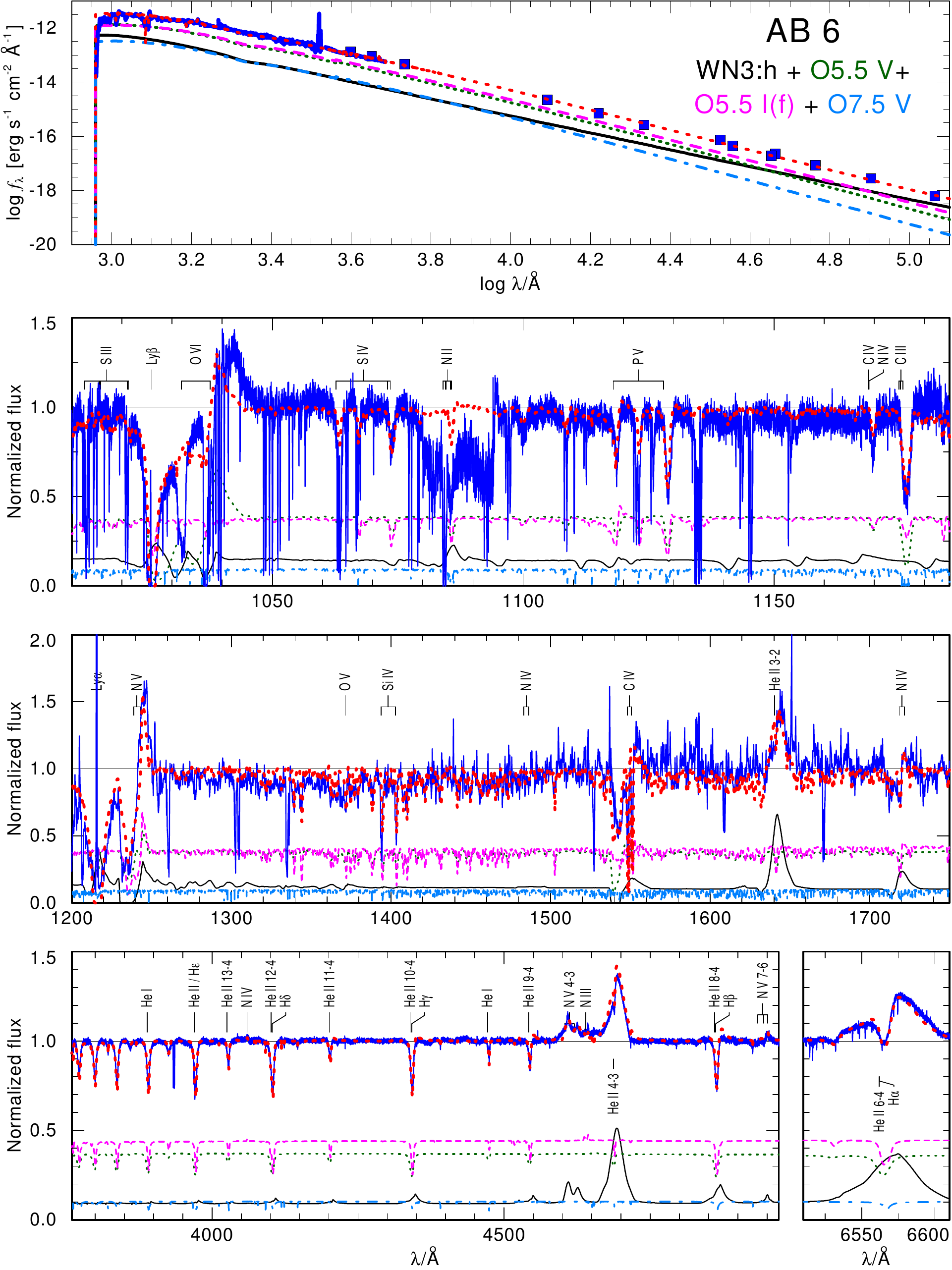}
  \caption{\emph{Upper panel:} Comparison between observed photometry and flux-calibrated IUE and FUSE spectra (blue squares and lines) with 
  the best-fitting model SED of AB\,6 (red dotted line), which is the sum of the reddened SEDs of stars A, B, C, and D (black solid, green dotted, pink dashed, and cyan dot-dashed lines). 
  \emph{Lower panels:} Comparison between normalized FUSE, IUE, and UVES spectra at $\phi{\approx}0.1$ and the best-fitting normalized spectrum 
  of AB\,6, comprising the weighted models of stars A, B, C, and D. Line styles are as in the upper panel.}
\label{fig:fit}
\end{figure*} 

Optically thin clumps are accounted for using the microclumping approach 
\citep{Hillier1984, Hamann1998},
where the population numbers are calculated in clumps that are a factor of $D$ denser 
than the equivalent smooth wind. We find $D{=}10$ provides a good agreement in all cases, and refrain from 
fitting this parameter. The stratification of clumps is not well 
constrained. Some studies suggest that clumping initiates slightly above the stellar surface at $r{\approx}1.1\,R_*$ because of line-driven instability \citep{Owocki1988, Feldmeier1997}, 
while other studies suggest that clumps already originate in subphotospheric layers \citep{Cantiello2009, Ramiaramanantsoa2018}. We assume $D{=}1$ at the photosphere, which reaches its
maximum value $D{=}10$ at $r{=}1.1\,R_*$. The detailed form of the stratification may have some effect on the line profiles, but does not significantly affect our results.
In this study, we ignore the effect of optically thick clumps or macroclumping \citep{Oskinova2007, Surlan2013}. While potentially important, high quality, phase dependent UV data 
are necessary to obtain constraints on the clump geometry. Overall, accounting for macroclumping can introduce an increase of up to 
a factor of ${\sqrt{D}{\approx}3}$ in the derived mass-loss rate \citep[e.g.,][]{Oskinova2007, Shenar2015}.

Because optical WR spectra  are dominated by recombination lines, whose strengths increase with $\dot{M} \sqrt{D}$, it is customary 
to parametrize their models using the so-called transformed radius,

\begin{equation}
 R_\text{t} = R_* \left[ \frac{v_\infty}{2500\,{\rm km}\,{\rm s}^{-1}\,}  \middle/  
 \frac{\dot{M} \sqrt{D}}{10^{-4}\,M_\odot\,{\rm yr}^{-1}}  \right]^{2/3}
\label{eq:Rt}
\end{equation}
 \citep{Schmutz1989}, defined so that equivalent widths of recombination lines of models  
with given abundances, $T_*$, and $R_\text{t}$ are 
approximately preserved, independent 
of $L$, $\dot{M}$, $D$, and $v_\infty$. 

In principle, the temperatures are derived from the ionization balance of different species, gravities from the strengths and shapes 
of Balmer lines, and wind parameters from wind lines (H$\alpha$, He\,{\sc ii}$\,\lambda 4686$, UV resonance lines). The projected rotational 
velocity $v_\text{eq} \sin i$ is derived by convolving the models with appropriate rotation profiles.
For the WR and O(f) components (stars A and C), 
we applied a 3D integration routine \citet{Shenar2014} to derive $v_\text{eq} \sin i$ from emission lines. 
We account for macroturbulence by convolving the model spectra with  
radial-tangential profiles with $v_{\rm mac}{=}20\,${\kms} \citep{Gray1975, SimonDiaz2007, Puls2008}.  
Abundances are derived from the strengths of corresponding spectral lines.
The synthetic composite spectrum is convolved with Gaussians to mimic the instrumental profiles of the observations. The models of stars A and B also account for Auger ionization via X-rays \citep{Baum1992}.

The total luminosity $\log L_\text{tot}$ and reddening $E_\text{$B{-}V$}$ are derived by fitting the total model flux to the observed SED. The reddening 
is modeled using a combination of reddening laws derived by \citet{Seaton1979} for the Galaxy and by \citet{Gordon2003} for the SMC, where $E_\text{$B{-}V$}{=}0.03$ 
is assumed for the Galaxy and the total-to-selective extinction is set to $R_\text{V}{=}3.1$.

The most challenging part is determining the light ratios simultaneously with the mass-loss rate of the WR star. 
This is because an increase of the mass-loss rate of the WR star can be compensated for by decreasing the contribution of the WR star to the total visual light, thereby 
inducing a stronger dilution of its lines.
The problem was solved iteratively, constraining first components whose light contribution is easier to establish.
Details regarding the analysis are given in Sects.\,\ref{subsec:specanstarD} - \ref{subsec:specanstarA}.

\begin{figure*}
\centering
  \includegraphics[width=\textwidth]{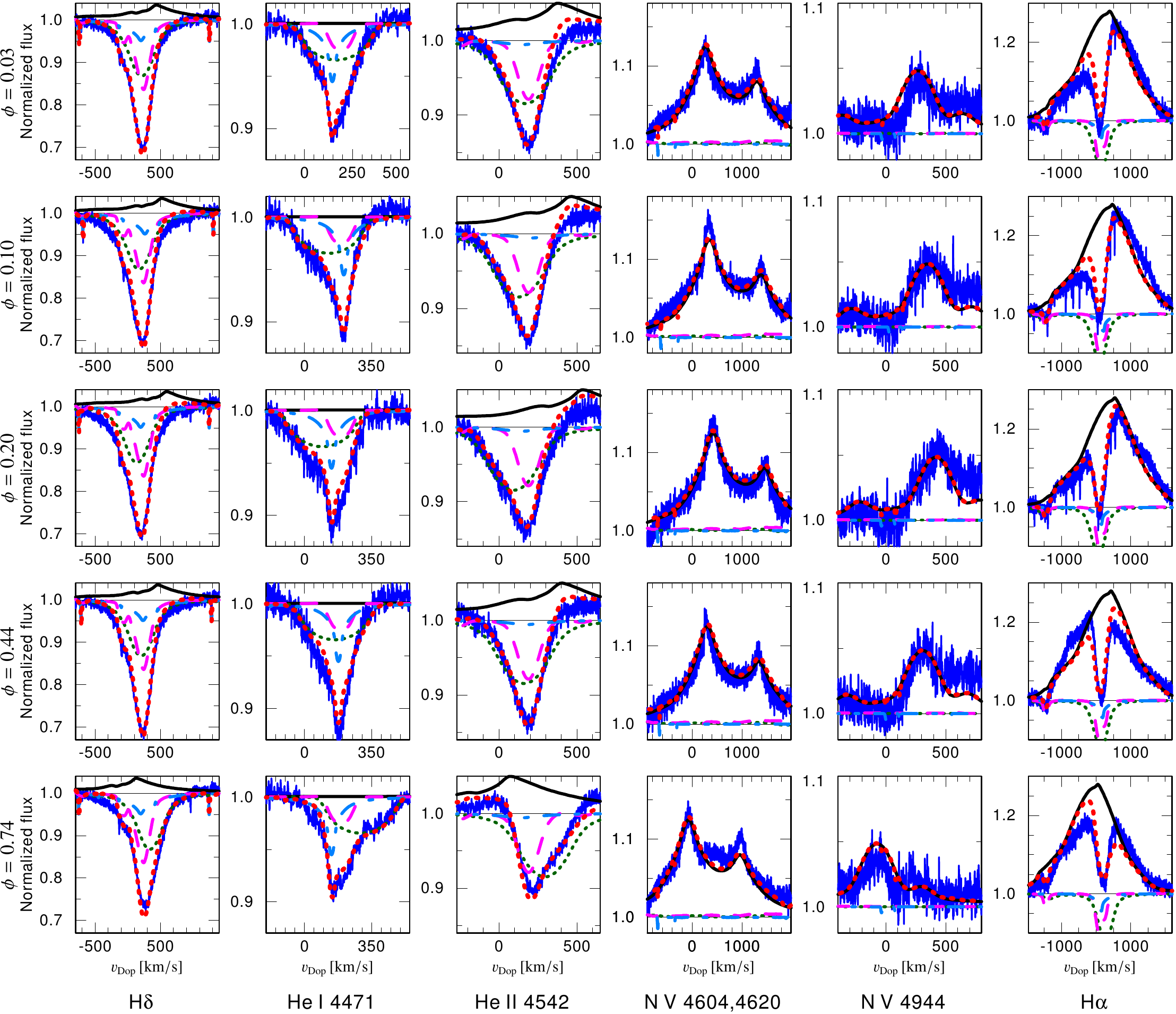}
  \caption{Phase-dependent fits for various phases and spectral lines  (see axes labeling). Line styles are as in 
  Fig.\,\ref{fig:fit}. 
}
\label{fig:phasefit}
\end{figure*}

The global 
fit to the data at $\phi{\approx}0.1$ is shown in Fig.\,\ref{fig:fit}.  A fit to specific lines at five different phases is shown 
in Fig.\,\ref{fig:phasefit}. The derived stellar parameters are given in Table\,\ref{tab:specan}, where we also give the  equatorial rotation velocity
$v_\text{eq}$, absolute visual 
magnitude $M_\text{V}$, light ratio in the $V$ band $f_\text{V} / f_\text{V, tot}$, and number of H-ionizing photons $\log Q$. Spectroscopic masses are calculated 
from $g_*$ and $R_*$. 
Errors on fundamental parameters 
follow from the sensitivity of the fit to their variation (see details below). The remaining errors are calculated via error propagation.

\subsection{Star D (O7.5~V)}
\label{subsec:specanstarD}

Despite its faintness, star D is the easiest to model. Given its relatively low temperature, 
it exhibits lines that are not present in any of the other stars, such as the Si\,{\sc iv}\,$\lambda \lambda 4089, 4116$ doublet, and weak 
He\,{\sc i} lines.  Moreover, its line profiles are very narrow and easy to distinguish from the other components (see Fig.\,\ref{fig:overview}). 
A weak signature of this star is also visible in strong He\,{\sc ii} lines, which enabled us to derive its temperature 
from the He\,{\sc ii}/{\sc i} balance and the ionization balance of N and O lines. 
A peculiarity of star D is its extremely narrow spectral lines, which could only be reproduced assuming $\xi_\text{ph}{\le}3$\,\kms 
and $v_\text{eq} \sin i{\le}3$\,kms. This result is reminiscent of the very low rotation and turbulence measured for the magnetic 
O9.7~V star \object{HD\,54879} \citep{Shenar2017b}, raising the possibility that star D may be a magnetic star.

With the temperature fixed, the light ratio of star D could be well constrained based on the strengths of its spectral lines, 
assuming no abnormal abundances, and is found to be 10\%.
Deriving $\log g_*$ for star D proved to be very difficult because the Balmer lines contain contributions from all components. 
Given its derived luminosity and its stellar parameters, star D is very likely a main sequence 
star, and so $\log g_*$ was taken with an appropriate value for late-type OB main sequence stars ($\approx 4.0\,$[cgs]). The value is consistent 
with the data within ${\approx}0.3$\,dex. 

The wind parameters of star\,D could not be constrained owing to its faintness. A terminal velocity of $v_\infty{=}2000\,$\kms\, is adopted 
based on scaling relations with the escape velocity \citep{Leitherer1992}. The mass-loss rate was fixed to \mbox{$\log \dot{M}{=}-9.4\,$[\myr]}, which is based 
on the typical mass-loss rates of late-type O stars \citep[e.g.,][]{Bouret2003, Marcolino2009, Shenar2017b} scaled down with $Z^{0.7}$ \citep{Vink2001}.

\subsection{Star B (O5.5~V)}
\label{subsec:specanstarB}

While all spectral lines of star B are entangled with the other components, their round profiles make star B easily recognizable 
in strong He lines (see Figs.\,\ref{fig:overview} and \ref{fig:phasefit}). The ionization balance of He was the main criterion 
for deriving $T_*$ for star B, which is found to be $41.5\,$kK.  We derive $v_\text{eq} \sin i{=}210\,$\kms, which is significantly higher 
than average \citep[e.g.,][]{Ramirez2013}, potentially due to past mass transfer
\citep[e.g.,][see Sect.\,\ref{subsec:evo}]{DeMink2013, Shara2017, Vanbeveren2017}. With these parameters 
fixed, the relative contribution of star B to the visual flux could be derived from the strength of the He lines assuming normal  
abundances. Star B is estimated to contribute 36\% in the visual.  The gravity cannot be accurately derived, but $\log g_*{=}4.0\,$\,\gcgs 
is consistent with the data within $0.3\,$dex, which suggests that star B is a main sequence star. The disentangled spectrum of star B is intermediate 
between O5 and O6, which led to the final estimate of O5.5~V.

\subsection{Star C (O5.5~I(f))}
\label{subsec:specanstarC}

Star C is most easily seen in apparently static N\,{\sc iii} and N\,{\sc iv} emission lines. However, it also 
contributes significantly to strong He\,{\sc i} and He\,{\sc ii} lines (see Fig.\,\ref{fig:phasefit}), which aids in constraining $T_*$. 
To obtain N\,{\sc iii} and N\,{\sc iv} in strong emission, $\log g_*$ are necessary. The value 
$v_\text{eq} \sin i$ could be derived based on both the weak emission lines of the star, which are formed very close to its photosphere, and from He\,{\sc i} absorption 
lines. The light ratio is determined from the overall strength of the He lines; the N lines are not helpful in this case because the N abundance 
of O(f) stars is uncertain. Star C is found to contribute roughly 45\% of the total visual light. After fixing its light ratio, the nitrogen abundance of star C could be 
constrained and is found to be ${\approx}20$ times the SMC average (see Table\,\ref{tab:specan}), which agrees with it being an evolved supergiant star. 
The disentangled spectrum of star C shows great similarity to the O5.5~I(f) star AzV 75 in the SMC 
\citep{Massey2009}, whose spectral type we therefore adopt.

\subsection{Star A (WN3:h)}
\label{subsec:specanstarA}

Finally, with the parameters of stars B, C, and D fixed, the parameters of star A can be derived. The previously derived light ratios 
imply that star A contributes about 9\% to the total visual light. With only N\,{\sc v} lines present among the nitrogen lines in the optical spectrum, 
we can conclude that the temperature of the WR star has to be larger than $75\,$kK. A good agreement is obtained for 80\,kK. In principle, $T_*$ can be arbitrarily larger, placing the star 
in the so-called $T_*{-}R_\text{t}$ degeneracy domain \citep{Hamann2006}. However, at significantly larger values of $T_*$ ($>90\,$kK),
the N\,{\sc v} doublet forms further out in the wind 
and the line profiles become broad and smeared unlike the sharply peaked profiles observed. Admittedly, the line profiles are sensitive to 
the adopted velocity law and clumping stratification. We therefore conservatively adopt a large upper bound on $T_*$. Correspondingly, $R_\text{t}$ has
a large lower bound. Nevertheless, the error on $\dot{M}$ remains relatively small because models of similar line strengths in the degeneracy domain 
maintain similar values of $\dot{M}$. 
A physical reason to prefer a lower $T_*$ is that higher temperatures generally require higher luminosities to reproduce the same observed flux, which 
only worsens the Eddington-limit problem. 

The gravity of the WR star has virtually no impact 
on its spectral appearance and is therefore not fitted here, but kept fixed to the value implied from the orbital mass and stellar radius of the WR star.
With the temperature and light ratio fixed, the mass-loss rate and detailed abundances immediately follow from the strengths 
of the emission lines in the spectrum.  Accounting for the four visible components implies that the luminosity of the WR star is smaller than previously derived by 
\cite{Shenar2016} and drops from $\log L{=}6.3$ to $5.9\,[L_\odot]$. 

The nitrogen abundance is found to be five times larger than what is expected from the CNO cycle equilibrium 
\citep[$X_\text{N}{=}0.0015$, e.g.,][]{Crowther2007, Hainich2015}. While the total error on $X_\text{N}$ 
may reach a factor of two due to uncertainties in the velocity law and clumping stratification, a mere inspection of the spectrum of star A in Fig.\,\ref{fig:disfit} suffices to 
indicate that it exhibits strong N\,{\sc v} lines. 
Larger-than-expected $X_\text{N}$ is reported for all SMC WN stars \citep{Shenar2016, Hainich2015}. 
For example, the WN3ha star AB\,12 was reported to have $X_\text{N}{=}0.009$ \citep{Hainich2015}. One way to enhance $X_\text{N}$ is by mixing carbon 
from the He-burning core into the H-burning shell, which is then converted to N during the CNO cycle \citep[e.g.,][]{Vincenzo2016}.
Such a process could be supported by tidally induced mixing \citep[e.g.,][]{Song2013}. Another possibility is that the original CNO abundance was larger than the SMC average. 
For example, the derived abundances of star C are suggestive of a total CNO abundance that is  ${\approx}5\%$ of the total stellar mass, i.e.,
three times larger than expected.
Generally, derived $X_\text{N}$ values for SMC WR stars show quite a spread, a fact 
which should be investigated in future studies.

\citet{Foellmi2003SMC} attributed the N\,{\sc iv}\,$\lambda 4060$ line to the WR component instead of star C. 
Star A seems to exhibit only N\,{\sc v} lines in the optical, which, following \citet{Smith1996}, implies the spectral type WN2h. However, as its spectrum is heavily 
diluted by the other components, we cannot exclude the presence of faint N\,{\sc iv} or C\,{\sc iv} lines associated with it. The disentangled
spectrum of star A, with its enhanced S/N of ${\approx}300$, is suggestive of a very faint and broad N\,{\sc iv} feature, but this could also arise from 
normalization issues. We therefore classify star A as WN3:h (``:'' stands for uncertain).


The simple $\beta$ law in Eq.\,(\ref{eq:betalaw}) did not provide a good fit to the sharply peaked profiles of the N\,{\sc v} lines 
(see Fig.\,\ref{fig:overview}). A better fit is obtained using an extension of the $\beta$ law, dubbed the
two-component $\beta$ law \citep{Hillier1999, Todt2015}, where a term similar 
to the standard $\beta$ law is added to Eq.\,(\ref{eq:betalaw}). We find that $\beta_1{=}1$ and $\beta_2{=}4$, which has a fractional 
contribution of $0.4$ for the $\beta{=}4$ component, provides a good fit to the N\,{\sc v} lines. Larger $\beta_2$ values of ${\approx}10$ \citep[e.g.,][]{Lepine1999} result in line profiles that are too narrow in all emission lines.
Future studies will try to consistently model the velocity field of the WR star in light of its known orbital mass and luminosity \citep{Sander2017}.
\subsection{Atmospheric eclipses in the UV}
\label{subsec:UVeclipse}

\begin{figure}[]
\centering
  \includegraphics[width=0.95\columnwidth]{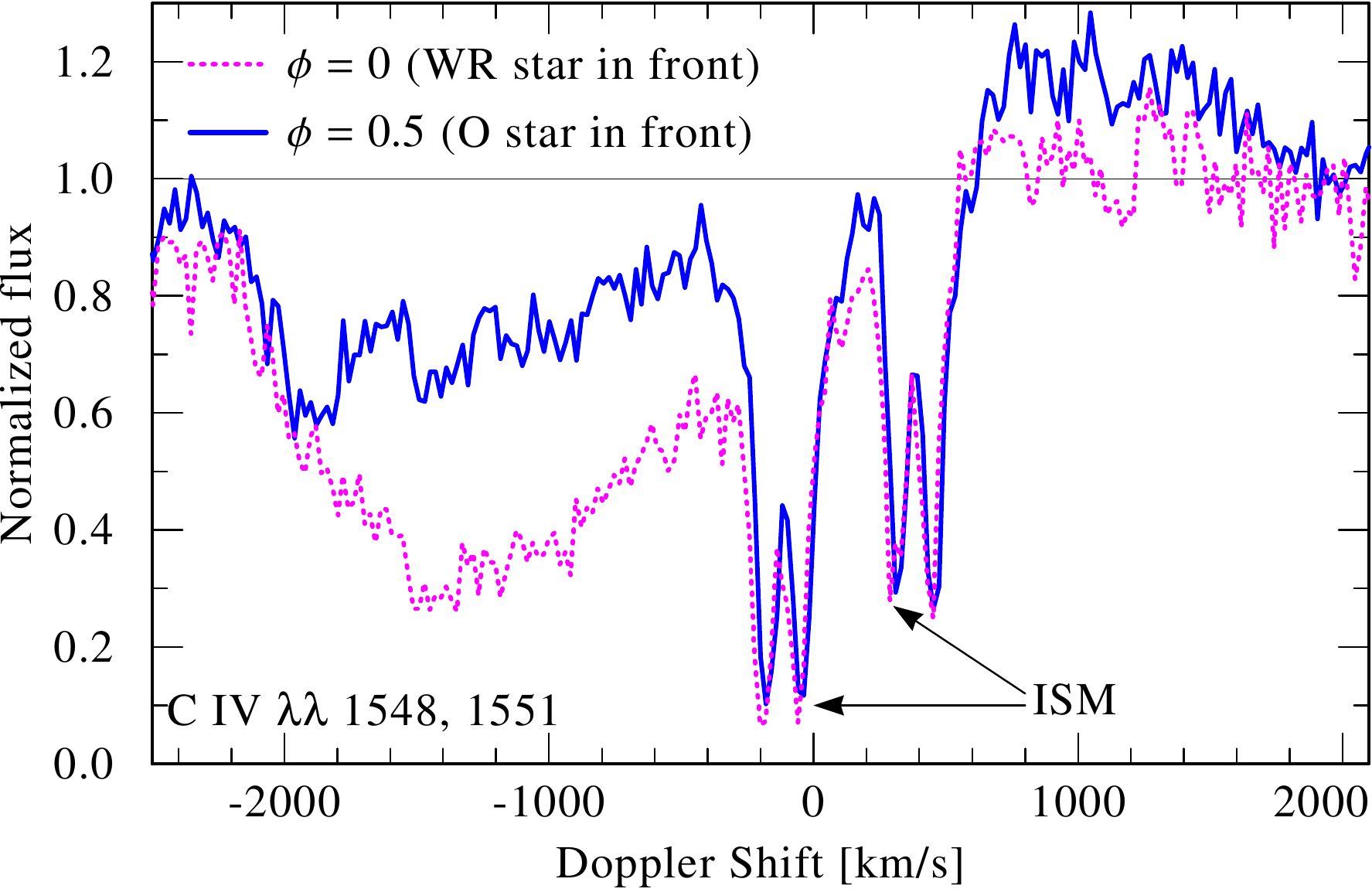}
  \caption{HST spectra at $\phi{\approx}0$ (pink line) and $0.5$ (blue line) in velocity space relative 
  to the blue component of the C\,{\sc iv} doublet corrected for the systemic velocity $V_0{=}196\,$\kms.}
\label{fig:UV}
\end{figure}

The terminal velocities are based in part on IUE and HST UV spectra available here. Two HST spectra of the C\,{\sc iv}\,$\lambda \lambda 1548, 1551$ resonance doublet 
taken close to inferior and superior WR 
conjunctions ($\phi{=} 0, 0.5$) are shown in Fig.\,\ref{fig:UV}. The spectra show significantly more absorption when the WR star is in front of the O star ($\phi{=}0$) 
compared to when the O star is in front of the WR star ($\phi{=}0.5$). This, as well as 
the behavior of the N\,{\sc v}\,$\lambda \lambda 1239,1243$ resonance doublet, was interpreted by 
\citet{Hutchings1993} as mutual irradiation effects of the winds of the WR star and its O companion. However, the latter authors were not aware of the two additional 
components in the system. In fact, the C\,{\sc iv} resonance doublet is very 
robust and is not sensitive to the ionization structure except for very extreme X-ray irradiation effects.


We wish to offer a somewhat simpler and, we believe, more plausible interpretation for the changing absorption strength of the C\,{\sc iv} resonance doublet. When the WR star is in front of the 
O star ($\phi{=}0$), the part of its wind that occults the O star leads to additional absorption along the line of sight toward the disk of the O star. This absorption 
occurs for Doppler shifts that lie within the minimum and maximum projected velocity of the wind that occults the O star. Given the orbital parameters, this should be 
slightly below the terminal velocity of the WR star. The absorption should occur both at redshifted 
and blueshifted wavelengths.  This is exactly what is observed in the HST spectra. 
An accurate modeling of this requires the calculation of the non-LTE radiative transfer in nonsymmetric geometry, which is beyond
the scope of this paper. With more UV observations covering the orbital phase, much better constraints on the mass loss of the primary and secondary could be obtained.

\section{Discussion}
\label{sec:disc}

\subsection{Eddington limit problem}
\label{subsec:eddi}

\begin{figure}[]
\centering
  \includegraphics[width=\columnwidth]{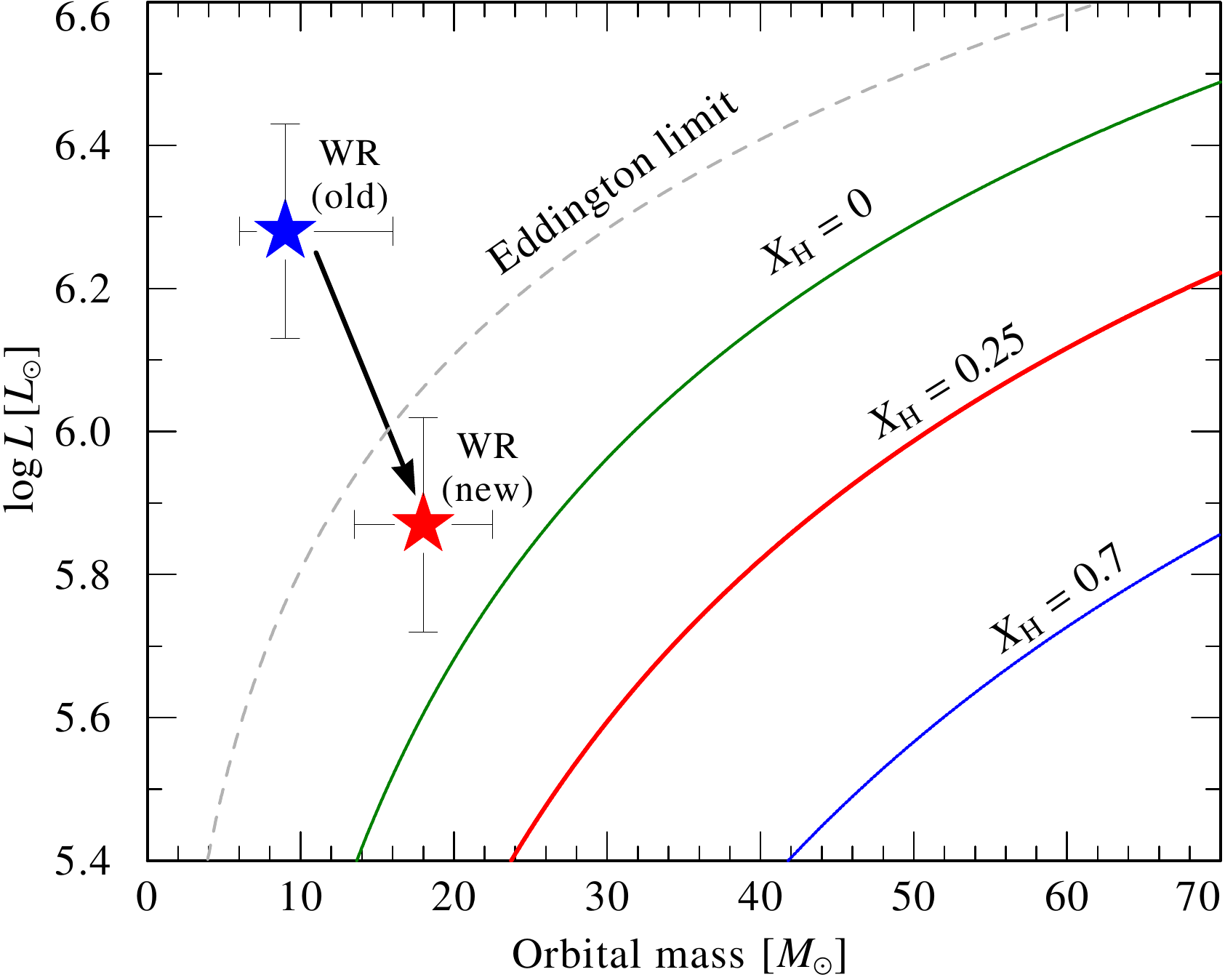}
  \caption{As in Fig.\,\ref{fig:Edin}, but including the new position of star A. Apart from the Eddington limit (dashed gray line), 
  we also plot mass-luminosity relations calculated for homogeneous stars with various hydrogen contents (see labeling).}
\label{fig:Edinnew}
\end{figure}

One of the primary motivations for our study was the fact that 
the WR primary in AB\,6 was found by \citet{Shenar2016} to exceed its expected luminosity by more than an order 
of magnitude, violating even the Eddington limit (see Fig.\,\ref{fig:Edin}). We now consider this problem 
in light of the new parameters derived in this study. In Fig.\,\ref{fig:Edinnew}, we plot 
a revised version of the $M-L$ diagram shown in Fig.\,\ref{fig:Edin}, but include the newly derived parameters for star A.
We also plot mass-luminosity relations calculated for homogeneous stars with different hydrogen contents by \citet{Graefener2011}.

Since $M_\text{WR}$ derived in this work is roughly a factor of two larger than the value reported by \citet{Foellmi2003SMC}, and the luminosity is about 0.4\,dex lower than the value 
reported in \citet{Shenar2016}, the WR primary is located below, yet close to, the Eddington limit (Fig.\,\ref{fig:Edinnew}), with an Eddington 
Gamma of $\Gamma_\text{Edd}{=}0.8$.
The urgent Eddington limit problem is therefore resolved via new, 
high quality UVES observations. 
The WR primary is still found to be overluminous compared to its respective homogeneous relation (with $X_{\rm H}{=}0.25$), implying that it is not homogeneous but 
rather core He burning and shell H burning. A similar result was obtained for most WR binaries in the SMC \citep{Shenar2016}. 

Considering the relatively
stable behavior of the WR component, its proximity to the Eddington limit is still surprising, especially 
when considering the full radiative force, which also accounts for line transitions. According to our model, 
the additional line acceleration in the deep quasi-hydrostatic layers because the Fe opacity peak 
causes the total outward force to exceed gravity, implying an unstable configuration. Similar results were obtained by \citet{Graefener2008}.
Given recent theoretical work on stellar envelope inflation in WR stars \citep[e.g.,][]{Graefener2011,Sanyal2015, Grassitelli2018}, it is possible that, while our model accurately describes  
the conditions in the wind of the WR star, it does not do so for its deep layers. However, this should have no bearing on our results.

\subsection{Evolutionary context}
\label{subsec:evo}

Since stars A${+}$B (WN3:h${+}$O5.5~V) constitute a close binary, they likely interacted in the past (and will probably do so in the future). Modeling their evolution 
therefore needs to account for binary interaction. Here, we use a grid of binary tracks calculated with the
BPASS\footnote{bpass.auckland.ac.nz} (Binary Population and Spectral Synthesis) stellar evolution code \citep{Eldridge2008, Eldridge2017} calculated for $Z{=}0.004$ to find a suitable evolutionary 
channel of the system. 

Each binary track is defined by a set of three parameters: the initial mass of the primary $M_{\rm i, 1}$, the initial mass ratio $q_\text{i}{=}M_\text{i,2} / M_\text{i,1}{<}1$, and the initial orbital
period $P_\text{i}$. The tracks  
are calculated at a spacing of $0.2$\,dex on $0{<}\log P\,[\text{d}]{<}4$,  a spacing of $0.2$ on $0{<}q_\text{i}{<}0.9$, 
and an uneven spacing of ${\approx}0.05-0.1\,$dex on $M_\text{i,1}$, amounting to roughly 6000 tracks. To find the best-fitting track and a corresponding age $t$, we minimize

\begin{figure}[]
\centering
  \includegraphics[width=\columnwidth]{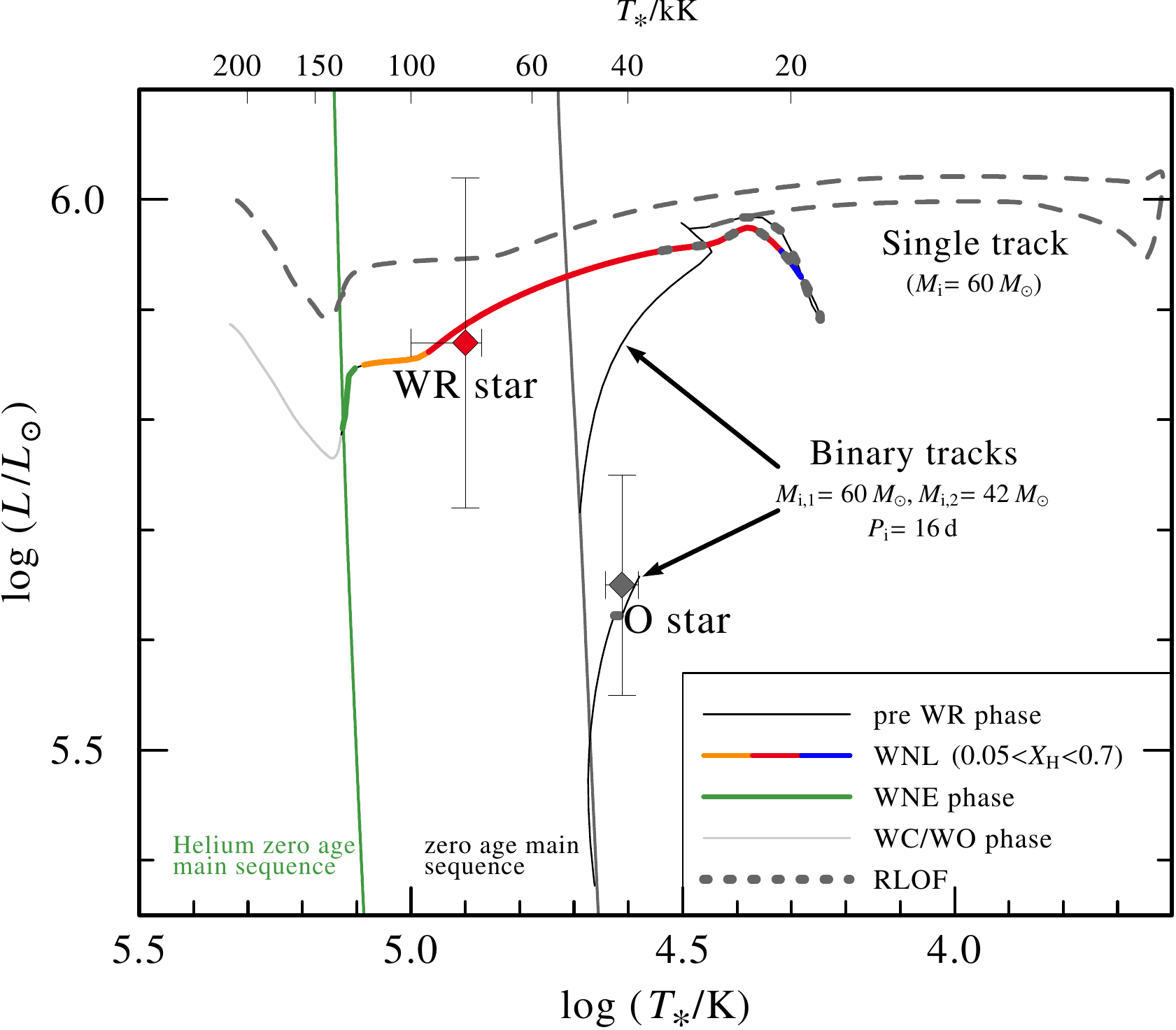}
  \caption{Best-fitting BPASS binary track for AB\,6, calculated for $M_\text{i}{=}60\,M_\odot$, $q_\text{i}{=}0.7,$ and $P_\text{i}{=}16\,$d. 
  The upper multi-colored  track depicts the evolution of the WR primary until its core-collapse. The colors refer to surface hydrogen mass fractions 
  of $0.45{<}X_\text{H}{<}0.7$ (blue), 
  $0.2{<}X_\text{H}{<}0.45$ (red), $0.05{<}X_\text{H}{<}0.2$ (orange), and $X_\text{H}{<}0.05$ (green). The WC and RLOF phases are also marked. 
  A track for a single $60\,M_\odot$ star is shown for comparison (dashed 
  gray line). 
  The lower track depicts the evolution track of the O-type companion until the core collapse of the primary.}
\label{fig:hrd}
\end{figure} 

\begin{figure}[]
\centering
  \includegraphics[width=\columnwidth]{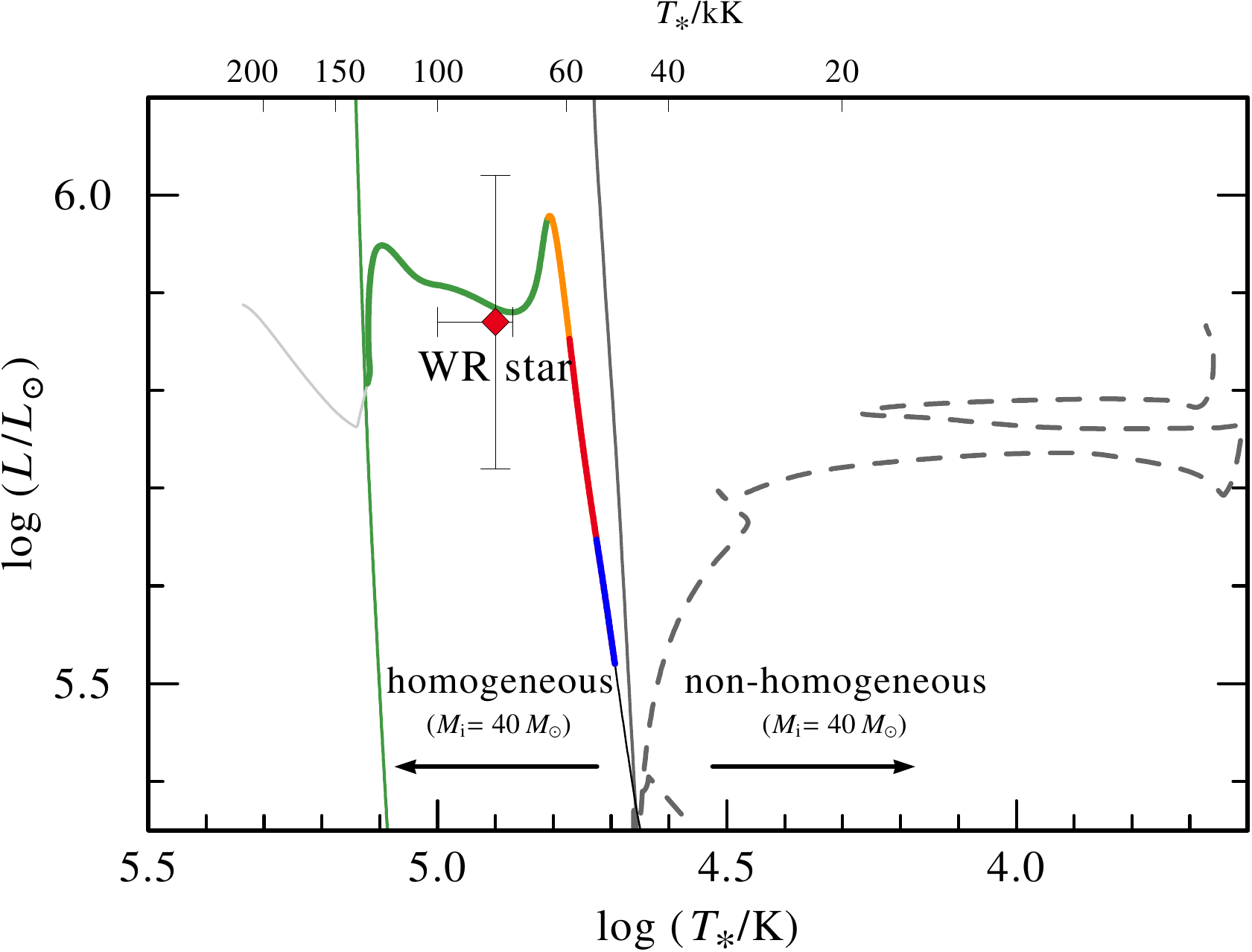}
  \caption{As in Fig.\,\ref{fig:hrd}, but depicting the best-fitting homogeneous evolution track calculated for a single $M_\text{i}{=}40\,M_\odot$ star. The nonhomogeneous 
  equivalent is shown for comparison (dashed line).
  }
\label{fig:hrd-hom}
\end{figure} 

\begin{equation}
 \chi^2\left(P_\text{i}, q_\text{i}, M_\text{i,\,1}, t\right) =\sum_{n=1}^{9} 
 \left(\frac{\text{O}_n - \text{E}_n\left(P_\text{i}, q_\text{i}, M_\text{i,1}, t\right)}{\sigma_n}\right)^2,
\label{eq:summin}
\end{equation}
where $\text{O}_n{\in}\left\{T_\text{WR}, \log L_\text{WR}, T_\text{O}, \log L_\text{O}, M_\text{orb,WR}, M_\text{orb,O}, \log P,\right.$
$\left. X_\text{H, WR}, q \right\}$ are the measured values for the
considered observables, and
$\text{E}_n\left(P_\text{i}, q_\text{i}, M_\text{i,1}, t\right)$ are the  
predictions of the evolutionary track defined by $P_\text{i}$, 
$q_\text{i}$, and $M_\text{i,1}$ at time $t$. $\sigma_n$ is defined by $\sigma_n{=}\sqrt{\Delta_n^2{+}\delta_n^2}$, where $\Delta_n$ is half the $n$'th parameter's 
grid spacing, and $\delta_n$ is the formal fitting error (see Tables\,\ref{tab:orbpar} and \ref{tab:specan}).

The best-fitting track is found to be that calculated with the initial parameters $M_\text{i, 1}{=}60\,M_\odot$, $q_\text{i}{=}0.7,$ and $P_\text{i} =16\,$d at an age of 3.9\,Myr. 
Roche-lobe overflow (RLOF) is predicted to have taken place shortly after core hydrogen exhaustion \citep[case B mass-transfer,][]{Kippenhahn1967} for  ${\approx}30$\,kyr, during which about $20\,M_\odot$ were lost from the primary 
and $5\,M_\odot$ were accreted by the secondary.
The tracks for the primary and secondary companions are shown in Fig.\,\ref{fig:hrd}. We also show a single-star track calculated by \citet{Eldridge2008} for an initial mass of 
$M_\text{i}{=}60\,M_\odot$ for comparison. In Table\,\ref{tab:evcomp}, we compare the observables $\text{O}_n$ to the prediction by the best-fitting binary track $E_\text{n}$, also stating  
the uncertainty $\sigma_n$ used in Eq.\,(\ref{eq:summin}). It is evident 
from Fig.\,\ref{fig:hrd} and Table\,\ref{tab:evcomp} that the binary track does very well in reproducing the observed parameters of the system. 

\setlength{\tabcolsep}{1.mm}
\begin{table}[]
\scriptsize
\small
\caption{Derived parameters for the WR binary ($O_\text{n}$) compared to predictions by the best-fitting binary track ($E_\text{n}$)}
\label{tab:evcomp}
\begin{center}
\begin{tabular}{l    c                   c                      c                       c               c               c                       c                       c                   c               }
\hline      
                          &       $P$           &      $M_\text{WR}$      & $M_\text{O}$   & $\log T_\text{WR}$  & $\log T_\text{O}$      & $\log L_\text{WR}$        & $\log L_\text{O}$          & $X_\text{H,WR}$    & $q$             \\ 
                          &       [d]          &   $[M_\odot]$           & $[M_\odot]$      & $[K]$            & $[K]$           & [\lum]        & [\lum]              &   -                & -                      \\                           
\hline                    
$O_\text{n}$               &      6.5          &      18                 & 41               &  4.90            &  4.61            &  5.87              &  5.65                &  0.25             &  2.2                   \\    
$E_\text{n}$               &      7.9          &      25                 & 45               &  4.90            &  4.61            &  5.89              &  5.63                &  0.21             &  1.8                     \\  
$\sigma_n$                &      1.5         &       8                  & 11               &  0.05            &  0.05            &  0.2               &  0.15                &  0.05             &  0.3                    \\
\hline   
\end{tabular}
\tablefoot{These best-fitting tracks are obtained for $M_\text{i,1}{=}60\,M_\odot, q_\text{i}{=}0.7$, and $P_\text{i}{=}16$\,d, and an age of 3.9\,Myr.
}
\end{center}
\end{table}

The tracks used in this work assume no initial rotation and do not perform a detailed evolution of the angular momentum of the system. Rapid initial rotation 
may cause the stars to undergo quasi-homogeneous evolution (QHE), avoiding expansion and, correspondingly, mass transfer in the system \citep{Meynet2003}. While studies point 
out that initial rotation velocities,  at least for O-type binaries, are unlikely to take near-critical values \citep{Ramirez2013, Ramirez2015}, 
this cannot be ruled out in specific cases.
For example, there is compelling evidence that the 
massive WR binary \object{SMC AB\,5} (\object{HD\,5980}) evolved via QHE \citep{Koenigsberger2014, Shenar2016}. 
To test this channel, we investigated a set of BPASS tracks calculated for homogeneous stars \citep{Eldridge2011} for a solution that best fits the observed properties of the WR component. 
The best-fitting track, found for $M_\text{i}{=}40\,M_\odot$, is shown in Fig.\,\ref{fig:hrd-hom}. This track 
never exceeds a radius of $9\,R_\odot$, remaining well within the current Roche radius of the WR star 
(see Table\,\ref{tab:orbpar}), which was likely not smaller in the past. Therefore, if QHE indeed characterizes the evolution of the system, the binary components would have avoided 
interaction. In this case, apart from tidal effects, their evolution can be modeled as for isolated single stars.

The best-fitting homogeneous track for the WR star  implies an age 
of 6.3\,Myr, which is roughly twice the age derived assuming binary evolution.
It is evident from Fig.\,\ref{fig:hrd-hom} that QHE cannot reproduce the hydrogen content of the WR star, with the track predicting $X_\text{H}{=}0$ at its 
observed HRD location.  The current stellar mass predicted by the track is $30\,M_\odot$, showing a larger deviation from our $18\,M_\odot$ measurement 
compared to the binary evolution track. However, the most severe issue is reproducing simultaneously the observed parameters of the O companion. To test whether QHE is consistent 
with the evolutionary status of the secondary, we use 
the BONNSAI tool.
Using $T_*, L, M_\text{orb}, v_\text{eq}$ derived for the secondary and the age derived for the WR star under the QHE assumption ($6.3\,$Myr), 
the algorithm tests whether a model exists that can reproduce the  properties 
of the secondary at a $95\%$ significance level. 
The results are conclusive: no solution can be obtained for the secondary. This result could be anticipated since the initial mass of the secondary would have to have been smaller than the  
$40\,M_\odot$ of the primary, although its current mass is estimated at $41\,M_\odot$. Moreover, after 6.3\,Myr, the O companion should be seen at an evolved stage, 
which is not observed. We can almost certainly reject QHE in the case of AB\,6. We conclude that the system exchanged mass in the past. 
This is supported by the higher-than-average rotation velocity found for star B, on par with Galactic candidates of spun up companions in WR binaries \citep{Shara2017}.

It is interesting to note that while our results suggest that the WR star in AB\,6 originally lost mass via RLOF, it is not clear that this mechanism was 
necessary for it to enter the WR phase. This can be easily seen in Fig.\,\ref{fig:hrd}, which shows that a single star with $M_\text{i}{=}60\,M_\odot$ is also predicted to become 
a WR star by the BPASS tracks. This prediction strongly relies on the mass-loss prescription adopted by the BPASS code. However, mass loss is still considered to be poorly 
understood, especially in light of stellar eruptions \citep[e.g.,][]{Smith2006}. 
A similar result was obtained in \citet{Shenar2016} for the other  
WR binaries in the SMC. This gives further evidence for the apparent lack of intermediate-mass WR stars that form via binary interaction (initial masses $10-40\,M_\odot$), 
which are expected to be abundant \citep{Bartzakos2001, Foellmi2003SMC}.

In the discussion of the evolution of the system, we fully ignored stars C and D. In principle, 
additional companions can affect the evolution of a binary \citep[e.g.,][]{Toonen2016}. However, these effects cannot be considered until 
more information is collected on stars C and D (see Sect.\,\ref{subsec:CD}).

\subsection{Nature of stars C and D}
\label{subsec:CD}

\begin{figure}[]
\centering
  \includegraphics[width=\columnwidth]{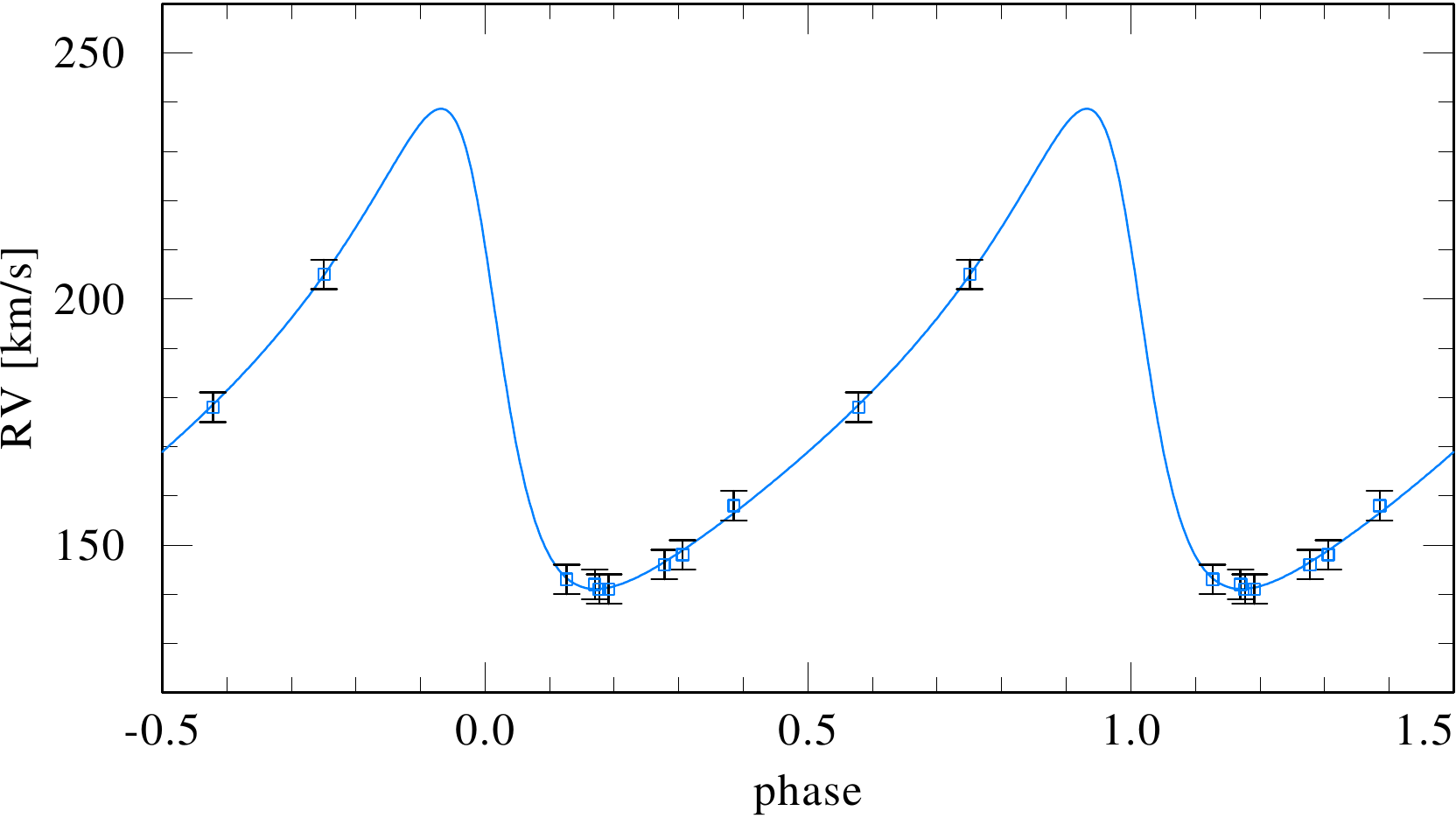}
  \caption{Preliminary SB1 orbital solution derived for star D
  }
\label{fig:orbD}
\end{figure}

A preliminary SB1 orbital fit 
to the measured RVs of start D is shown in Fig.\,\ref{fig:orbD}. The 
resulting parameters are $P{=}139.1\,$d, $V_0{=}181\,$\kms, $K{=}49$\,\kms, $\omega{=}64^\circ$, $e{=}0.46$, and $T_0{=}51937.3$\,[MJD]. For meaningful 
errors, more observations are necessary,  preferably taken near periastron. 
Regardless, it is clear that star D orbits a massive object with a period on the order of $100\,$d.

The first possibility is that star D orbits the WR binary, forming an hierarchical triple system. Adopting the spectroscopic mass derived for star D, 
the period derived above and the masses of stars A and B imply a separation of ${\approx}500{\pm}50\,R_\odot$ between star D and the WR binary. Is such an orbit stable? 
Analytical stability criteria by \citet{Mardling1999} suggested that, for the orbital parameters derived above, the orbit would be unstable for separations smaller than $450{\pm}100\,R_\odot$
\citep[cf.\ Eq.\,23 in][]{Toonen2016}. The stability condition is therefore only marginally fulfilled, 
and while we cannot rule out that star D revolves around the WR binary in a 140\,d period, this configuration seems unlikely.


\begin{figure}[]
\centering
  \includegraphics[width=\columnwidth]{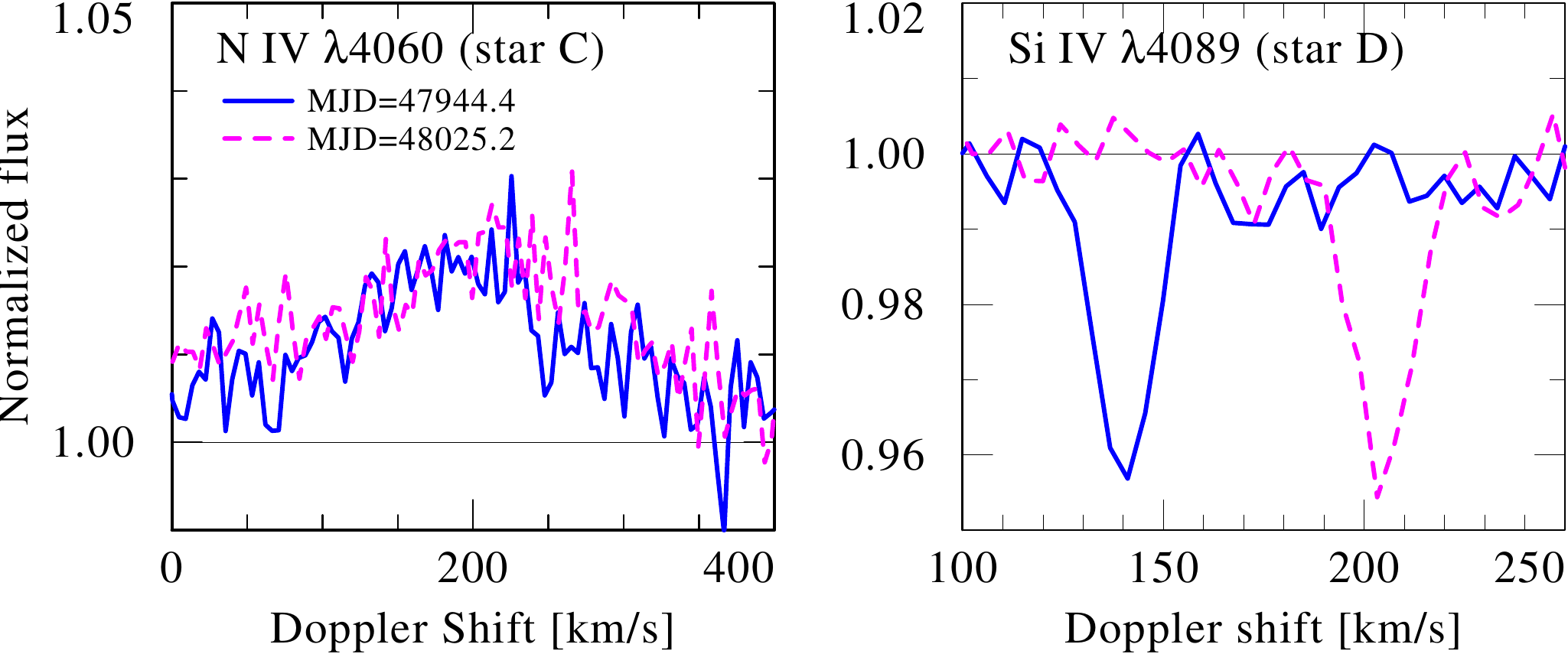}
  \caption{Two UVES observations (MJD{=}57944.4 58025.2 -  57944.4) that correspond to the maximum RV of star D. The left panel shows the N\,{\sc iv}\,$\lambda 4060$ line that 
  belongs to star C, and the right panel shows the Si\,{\sc iv}\,$\lambda 4089$ line that belongs to star D. No antiphase motion can be seen 
  in the spectrum of star C.
  }
\label{fig:starsCD}
\end{figure}

The only alternative is that star D forms a second binary.
The most obvious candidate for a companion to star D would be star C. However, the RVs derived for star C are not consistent with it forming a binary with star C. 
We cannot detect any significant RV variability for star C within the $10\,${\kms} error, and no antiphase trend is seen for the derived values 
(cf.\ Table\,\ref{tab:RVs} and Fig.\,\ref{fig:RVs_all}). 
Fig.\,\ref{fig:starsCD} illustrates the long-term motion of star D in the Si\,{\sc iv}\,$\lambda 4089$ line, 
whose RV changes by ${\approx 65\,}${\kms} within ${\approx}80\,$d.  For comparison, we also 
plot the N\,{\sc iv}\,$\lambda 4060$ line that originates in star C.
No antiphase motion relative to star 
D is seen. If the companion is truly star C, we would expect an antiphase RV change whose amplitude scales 
with the mass ratio $M_\text{D} / M_\text{C}$. Assuming $M_\text{D}{\gtrsim}20\,M_\odot$ and nondetectability for an RV change smaller than 
10\,{\kms}, the mass of star C would have to be

\begin{equation}
 \label{eq:massmin}
 M_\text{C}{=}\frac{\text{RV}_\text{D}}{\text{RV}_\text{C}} M_\text{D} > \frac{65}{10} \cdot 20{=}130\,M_\odot,
\end{equation}
which is very high and, while not impossible for an O(f) star, is not consistent with the luminosity and spectroscopic mass derived in this study for star C.


The preliminary orbital parameters for star D imply a mass function of $f{=}1.3\,M_\odot$. Denoting the companion of star D with the letter X, a lower limit 
for the mass of the companion of star D can be obtained by solving the following cubic inequality:

\begin{equation}
 f \equiv \frac{ M^3_\text{D} \sin^3 i } { \left(M_\text{X} + M_\text{D}\right)^2}{<}\frac{M^3_\text{D}}{\left(M_\text{X} + M_\text{D}\right)^2},
 \label{eq:lowlim}
\end{equation}
which yields $M_\text{X} > 11\,M_\odot$.
Hence, the companion of star D has to be a massive star (earlier than B3~V), for which there are  
three alternatives. The first option is that the companion of star D is star C and that $M_\text{C} > 130\,M_\odot$ (see Eq.\,\ref{eq:massmin}). However, this is not 
favored, as argued above. 
The second option is that the companion of star D is a very faint early B-type star. A B2 dwarf would fulfill $M > 11\,M_\odot$ and would have an absolute 
visual magnitude of $M_\text{V} {\approx} -2.5\,$mag \citep{Schmidt-Kaler1982}, which may  fall below our detection 
limit. The third option is that the companion is a massive black hole (BH). 
Using the parameters of star D from Table\,\ref{tab:specan} and adopting the minimum mass of $11\,M_\odot$ for a hypothetical BH on a circular orbit, 
Bondi-Hoyle accretion \citep{Bondi1994} would imply an X-ray luminosity of ${\sim}10^{31}$\,erg\,s$^{-1}$, which is negligible compared 
to WWC X-ray emission \citep{Guerrero2008}. Hence, the presence of a BH cannot be ruled out by
existing X-ray observations of this system.

To conclude, we suggest that the most likely configuration of AB\,6 is a quintuple system, in which stars A${+}$B form the 6.5\,d WR binary, star C a single star, 
and star D forms a second binary with either a BH or a B2~V star (Fig.\,\ref{fig:quadquint}). 
It remains unclear whether stars C, D, and the putative companion of star D are gravitationally bound to the A+B WR binary. Our results suggest that 
$V_{0, \text{A+B}} = 196{\pm}4\,$\kms, $V_\text{0, C} = 211{\pm}4\,$\kms, and $V_\text{0, D+X} = 181{\pm}9\,$\kms, although these errors (especially of stars C and D) may be well 
underestimated owing to systematics. Considering the errors, a gravitationally bound quintuple system can be neither confirmed nor ruled out. 
Since AB\,6 is seen as a point source in our UVES exposure at a seeing of $\text{FWHM}{=}1.4''$, and assuming no (extremely unlikely) line-of-sight contamination,
we can estimate that the five components should be confined within 
a circle with a radius of ${\approx}0.2\,$pc at the SMC distance.
To know whether the system is gravitationally bound, more spectra of AB\,6 are needed to further constrain the orbit of star D, and high resolution 
images of the immediate surroundings of AB\,6  (e.g., HST, adaptive optics) may help to spatially resolve this important system. 


\section{Summary}
\label{sec:summary}

This study presented new, high quality UVES observations
of the shortest period SMC WR binary, SMC AB\,6. The very low orbital mass and high luminosity of the 
WR component reported in the past suggested that it exceeds the Eddington limit \citep{Shenar2016}. Our study aimed at understanding these peculiar results and investigating 
the evolutionary history of this system. To achieve this, we performed an orbital, light-curve, and spectral analysis, and compared our results  to evolutionary tracks 
calculated with the BPASS code.
We conclude the following:

\begin{itemize}
 \item \emph{True members of the WR binary resolved:} The WR binary comprises an WN3:h component (star A) and a relatively rapidly rotating 
($v_\text{eq}{=}265\,$\kms) O5.5~V companion (star B), orbiting each other at a period of 6.5\,d. The new orbit derived implies $M_\text{WR}{=}18\,M_\odot$, which is twice the value reported in the past. 

 \item \emph{AB\,6 is a quadruple or quintuple system:}  The presence of an emission-line O5.5~I(f) star (star C) and of a narrow-line, fainter O7.5~V star (star D) 
can be clearly seen in the spectrum. The RVs of star D suggest that it belongs to a second binary. The lack of opposite 
RVs for star C suggests that star D orbits a BH or a B2~V star.  To know whether these additional components are gravitationally bound to 
the WR binary, more observations are needed.

 \item \emph{Eddington limit puzzle resolved:} The newly derived luminosity ($\log L{=}5.9\,[L_\odot]$) and orbital mass ($18\,M_\odot$) of the WR component 
 no longer place it above the Eddington limit, although it remains significantly overluminous compared to a homogeneous star with the same hydrogen content.
 
 \item \emph{Evidence for mass transfer:} The properties of the WR binary can be well reproduced 
 assuming that this binary originates in a system with initial masses $M_\text{i,1,2}{=}60,40\,M_\odot$ and 
an initial period of $P_\text{i}{\approx}16\,$d. The system very likely experienced mass transfer via RLOF in the past, with about $20\,M_\odot$ lost from the primary and 
$5\,M_\odot$ accreted by star B during RLOF. 
This is supported by the relatively rapid rotation  of star B. The age of the system is estimated at 3.9\,Myr. However, according to the same set of evolution tracks, any $60\,M_\odot$
star would eventually reach the WR phase, regardless of binary interaction.
\end{itemize}

It appears that while the new UVES observations have answered decade-long questions related to this unique and important system, 
they also create new questions. How many more components will we find with new data? This study comes to show that our understanding of massive stars 
is strongly dependent on our capability of resolving them and coping with their multiplicity. We encourage new observations of this system to understand the nature of stars C and D, and their 
potential role in the evolutionary history of this unique WR multiple system.

\begin{acknowledgements}
We thank our referee, Paul Crowther, for helping to improve our manuscript.
Based on observations collected at the European Organization for Astronomical Research in the Southern Hemisphere under ESO programme 099.D-0766(A) (P.I.: T.
Shenar). T.S. acknowledges
support from the German “Verbundforschung” (DLR) grant 50 OR 1612. V.R. is grateful for financial support from Deutscher Akademischer Austauschdienst (DAAD), as
a part of Graduate School Scholarship Program. 
A.M. is grateful for financial aid from NSERC (Canada) and FQRNT (Quebec).
LMO acknowledges support by the DLR grant 50 OR 1508. A.S. is supported by the Deutsche Forschungsgemeinschaft
(DFG) under grant HA 1455/26. This research made use of the VizieR catalog access tool, CDS, Strasbourg, France. The original description of the VizieR service
was published in A\&AS 143, 23. Some data presented in this paper were retrieved from the Mikulski Archive for Space Telescopes (MAST). STScI is operated by the
Association of Universities for Research in Astronomy, Inc., under NASA contract NAS5-26555. Support for MAST for non-HST data is provided by the NASA Office of
Space Science via grant NNX09AF08G and by other grants and contracts. 
\end{acknowledgements}
\bibliography{literature}
%


\begin{appendix}
%
%
%

%

\end{appendix} 
\end{document}